\documentclass[12pt,preprint]{aastex}
\shorttitle{RCS-1 spectroscopy}
\shortauthors{Gilbank et al.}

\newcommand\gsim{\gtrsim}
\newcommand\lsim{\lesssim}
\newcommand\kms{km s$^{-1}$}
\newcommand\bgc{B$_{gcR}$}

\begin{document}

\title{Spectroscopy of moderately high-redshift RCS-1 clusters\footnote{This paper includes data gathered with the 6.5 meter Magellan Telescopes located at Las Campanas Observatory, Chile.}~\footnote{This work is based on observations obtained at the Gemini Observatory, which is operated by the Association of Universities for Research in Astronomy, Inc., under a cooperative agreement with the NSF on behalf of the Gemini partnership: the National Science Foundation (United States), the Particle Physics and Astronomy Research Council (United Kingdom), the National Research Council (Canada), CONICYT (Chile), the Australian Research Council (Australia), CNPq (Brazil) and CONICET (Argentina).}}

\author{David G. Gilbank and H. K. C. Yee}
\affil{Department of Astronomy and Astrophysics, University of Toronto,
  50 St George Street, Toronto, Ontario, Canada, M5S 3H4}
\email{gilbank@astro.utoronto.ca, hyee@astro.utoronto.ca}

\author{E. Ellingson}
\affil{Center for Astrophysics and Space Astronomy, University of Colorado at Boulder, CB389, Boulder, CO 80309}
\email{Erica.Ellingson@colorado.edu}

\author{M. D. Gladders\altaffilmark{*}}
\affil{Department of Astronomy and Astrophysics, 
University of Chicago, 5640 S. Ellis Ave., Chicago, IL, 60637}
\email{gladders@oddjob.uchicago.edu}

\author{L. F. Barrientos}
\affil{Departamento de Astronom\'{\i}a y Astrof\'{\i}sica,
Universidad Cat\'{o}lica de Chile, Avenida Vicu\~na Mackenna 4860, Casilla 306, Santiago 22, Chile}
\email{barrientos@astro.puc.cl}

\and

\author{K. Blindert}
\affil{Department of Astronomy and Astrophysics, University of Toronto,
  50 St George Street, Toronto, Ontario, Canada, M5S 3H4}
\email{blindert@astro.utoronto.ca}

\altaffiltext{*}{Visiting Associate, The Observatories of the Carnegie Institution of
Washington, 813 Santa Barbara St., Pasadena, CA 91101}

\begin{abstract}
We present spectroscopic observations of 11 moderately high-redshift (z$\sim$0.7--1.0) clusters from the first Red-Sequence Cluster Survey (RCS-1).  We confirm that at least 10 of the 11 systems represent genuine overdensities in redshift space and show that for the remaining system, the spectroscopy was not deep enough to confirm a cluster.   This is in good agreement with the estimated false positive rate of $<$5\% at these redshifts from simulations. We find excellent agreement between the red-sequence estimated redshift and the spectroscopic redshift, with a scatter of 10\% at z$>$0.7. At the high-redshift end (z$\gsim$0.9) of the sample, we find two of the systems selected are projections of pairs of comparably rich systems, with red-sequences too close to discriminate in $(R-z^\prime)$ colour.  In one of these systems, the two components are close enough to be physically associated.  For a subsample of clusters with sufficient spectroscopic members, we examine the correlation between \bgc~(optical richness) and the dynamical mass inferred from the velocity dispersion.  We find these measurements to be compatible, within the relatively large uncertainties, with the correlation established at lower redshift for the X-ray selected CNOC1 clusters and also for a lower redshift sample of RCS-1 clusters.  Confirmation of this and calibration of the scatter in the relation will require larger samples of clusters at these and higher redshifts.
\end{abstract}


\section{Introduction}
Clusters of galaxies provide probes of cosmological parameters, such as
those describing the equation of state of dark energy, and are laboratories for
studying galaxy evolution.  In order to place the strongest constraints
on cosmological parameters, clusters at redshifts as high as z$\sim$1
are crucial \cite[e.g.,][]{2002ApJ...577..569L, Lima:2004jt}.  Observations at these
redshifts also provide vital insight into the evolution of galaxies, at an epoch when clusters appear to be assembling (e.g., \citealt{2004astro.ph..8165F} and references therein).

Previously, only a handful of systems were known at such redshifts.
These were selected in a variety of ways, e.g., from an optical
photographic survey \citep{gho86} and X-ray selection
(e.g., \citealt{1999AJ....118...76R}).  With the advent of the Red-Sequence
Cluster Survey \citep[RCS-1,][]{Gladders:2005oi} the size of the sample of
clusters at these redshifts has increased manyfold.  More importantly,
this larger sample possesses a homogenous and readily quantifiable
selection function \citep{Gladders:2002ui}.  

RCS-1 is a 90 square degree optical survey (72 square degrees after cutting to the highest photometric data quality) aimed at finding galaxy clusters out to redshifts of order unity using only moderate-depth $R$- and $z^\prime$-band imaging.  The primary science goal of the survey is to measure cosmological parameters through the evolution of the cluster mass function \citep{Gladders:2006pp}.  In order to do this efficiently, the survey data themselves are used to estimate the redshift and the mass of the clusters.  The redshift is estimated via the position of the cluster red-sequence \citep{Gladders:2006pp} and the mass proxy used is optical richness as measured by the B$_{\rm gc}$ parameter (see \citealt{Yee:2003we} and references therein)\footnote{We use a modified version of the B$_{\rm gc}$ parameter \citep[\bgc, see][]{Gladders:2005oi}, considering only galaxies with colours compatible with the red-sequence at the estimated redshift of the cluster}.  

In this paper, we present 8-m class spectroscopic observations of a subsample of 11 moderately high-redshift RCS clusters in order to confirm the reality of these systems, the accuracy of the redshift estimate and the applicability \bgc~as a mass estimator.

\section{Observations \& data reduction}

\subsection{Sample selection}
Cluster candidates were selected from a preliminary version of the RCS-1 cluster catalogue, before the photometric calibration was finalised.  The selection was designed to be as close as possible to a richness-selected sample within the desired redshift range and available RA range. The earliest cluster candidates for follow-up were prioritised by visual inspection.  Possible biases associated with this selection are discussed in \S4.3.  Recalibration of the photometry affects the estimated redshift, the measured richness and the detection significance of a cluster.  As a result, two
clusters do not appear in the final catalogue (see \citealt{Gladders:2005oi} for details of two patches).  Both of these were rejected due to the strict significance threshold of 3.3-$\sigma$ (equivalent to
a probability of only 1 in 1000 of occurring by chance) used in the final catalogue.  One cluster appeared in an early preliminary (December 2004) catalogue  and its properties (estimated redshift, significance, richness) from this were used.  In order to measure the parameters of
the other candidate in a consistent way with those of the final catalogue,
the RCS cluster-finding algorithm was re-run with a lower threshold
cutoff.  It was only necessary to lower the threshold to 3.2-$\sigma$
in order to recover the remaining candidate.  Cluster candidate
parameters quoted in this paper are taken from an improved later generation (December 2005) catalogue.

\subsection{Spectroscopic observations}
Spectroscopy was carried out in three runs on the 6.5-m Walter Baade
Telescope: 2001 December 11-13 and 2002 January 15-16 using the Low
Dispersion Survey Spectrograph 2 \cite[LDSS-2,][]{Allington-Smith:1994qp}; December 2003 with the Inamori Magellan Areal Camera and Spectrograph \citep[IMACS,][]{Bigelow:1998fj}; and in two queue runs (programme IDs GN-2002A-Q42 and GN-2003B-Q-19) with the Gemini Multi-Object Spectrograph (GMOS, Hook et al.\ 2004) on Gemini North. The observations are listed in Table 1.  LDSS-2 used the {\sc
  med/red} grism giving a dispersion of 5.3\AA/pixel centred around
5500\AA\ with a nominal resolution of 13.3\AA\ over a $\sim$6.5 $\times$ 5 arcmin field. The masks comprised $\sim 30$ 7-10 arcsec long and typically 1 arcsec wide slits, observed
for the total integration times listed in Table 1, usually split into 20
minute sub-exposures. For IMACS, the G200 grism was used, giving a dispersion of
2.0\AA/pixel centred around 6600\AA\ with a resolution of 11.0\AA\ over a 27 arcmin diameter field.  The
IMACS masks consisted of $\sim$150 slits, and the
instrument was used in nod \& shuffle (N\&S, e.g.,
\citealt{2001PASP..113..197G}) mode.  Exposures of 60s were
taken, and after every exposure the telescope pointing was 'nodded' 1.4 arcsec along the slit, and the charge shuffled along the detector. This procedure was repeated for the total exposure times given in Table 1.  When
the data were read out, this resulted in two observations of the same object:
with observations of the night sky spectrum in the second exposure at
the position of the object in the first exposure and vice-versa.  2D
sky subtraction could then be accomplished by simply subtracting one shuffled
region from the other, producing a positive object spectrum at the
first position and a negative spectrum of the object at the nodded
position.  The GMOS$-$N observations used the R150 grism and the detector was binned 2$\times$2 giving a resolution of 11.4\AA\ at a dispersion of 3.5\AA/pixel over a 5.5 arcmin field.  RCS1417$+$5305 was observed in nod \& shuffle mode, whereas RCS1620$+$2929 was observed as classical MOS.

\begin{deluxetable}{lc}
\tablecaption{Summary of integration times for each cluster.  
  }
\tablewidth{0pt}
\tablehead{
\colhead{Mask name} & \colhead{Total exposure time (ks)} 
}
\startdata
\cutinhead{LDSS-2 (classical MOS)}
RCS033414-2824.6A & 3.60\cr
RCS033414-2824.6B & 3.60\cr
RCS034850-1017.6A & 5.40\cr
RCS043938-2904.8A & 14.65\cr 
RCS043938-2904.8B & 7.20\cr
RCS044111-2858.3A & 10.80\cr
RCS110246-0426.9B & 5.72\cr
RCS110634-0408.9A & 5.40\cr
RCS110708-0355.3A & 8.10\cr
RCS110723-0523.2A & 3.00\cr
RCS110723-0523.2B & 1.50\cr
\cutinhead{IMACS (nod \& shuffle)}
RCS035231-1020.7  & 1.5\cr
RCS043938-2904.8  & 5.13\cr
\cutinhead{GMOS-N (nod \& shuffle)}
RCS141658+5305.2A & 7.68\cr
RCS141658+5305.2B & 21.1\cr
\cutinhead{GMOS-N (classical MOS)}
RCS162009+2929.4A & 9.00\cr
RCS162009+2929.4B & 6.18\cr
\hline
\enddata
\end{deluxetable}

\subsection{LDSS-2 reduction}

The LDSS-2 reduction was performed using a set of {\sc python} routines written by Dan
Kelson and available from {\tt http://www.ociw.edu/$^\sim$kelson/}.
This software is based on earlier {\sc fortran} routines whose
operations are detailed in \citet{2000ApJ...531..159K}.  The approach
used was to compute the y-(spatial-)distortion along the slits by
measuring the curvature of slit edges in a flatfield using {\sc
  getrect}.  Slit edges were identified automatically from the
flatfields with the {\sc findslits} task and in a few cases adjusted
manually using {\sc editslits}. The x-(spectral-)distortion was
calculated by tracing lines from the arc lamp for each slit.  The
wavelength calibration was then calculated automatically ({\sc
  waverect}) from the lamp using a list of reference wavelengths and
their approximate relative intensities, coupled with estimates of the
starting and ending wavelength and approximate dispersion for each
dataset. A zero-order shift of the wavelength calibration was then
calculated using the night sky lines in the science data and applied,
if necessary, to compensate for flexure in the instrument.  The
measured distortions were then used to correct the flatfields, which
were used to normalise the spectra. The x-, y-, and wavelength
distortions calculated were used to resample the 2D spectral data to a
linear frame with the spatial and spectral distortions removed and all
the slits aligned in wavelength space.  This was performed in a single
operation using {\sc unrect}. We then ran the {\sc iraf}\footnote{{\sc
    iraf} is distributed by the National Optical Astronomy Observatory
  which is operated by AURA Inc.\ under contract with the NSF.} task
{\sc apall} on these 2D rectified data to extract and sky-subtract the
spectra.

\subsection{IMACS reduction}
The IMACS data were reduced using an early version (1.02) of the
Carnegie Observatories System for MultiObject Spectroscopy ({\sc
  COSMOS}) software\footnote{{\tt
    http://llama.lco.cl/$^\sim$oemler/COSMOS.html}} written by Gus
Oemler.  This uses a map of the IMACS instrumental distortions to
enable accurate rectification of the spectra.  After checking the
alignment of the mask to the sky using the {\sc apertures} task on a
direct image through the mask, alignment of the spectral mask was
performed by running the {\sc align-mask} task on a calibration arc.
This task fits for shift and rotation of the mask by comparing the
predicted positions (using the distortion model) of a few bright lines
in the arc with their observed positions. Once the low order
(alignment) terms have been fitted, the mapping between CCD detector
coordinates and spectral coordinates (wavelength and slit position) for
the comparison arc was calculated using {\sc map-spectra}.  This
mapping was tweaked to fit out the higher order residuals with {\sc
  adjust-map} through comparison of the predicted positions of the full
list of lines in the arc with their measured positions. After checking
the mapping by overplotting the lamp line positions on the arc image,
the mapping was applied to the spectroscopic flatfield (with {\sc
  Sflat}).  Once all the mappings had been calculated, the science
frames were debiased and flatfielded using the {\sc biasflat} routine,
and {\sc extract-2D} used to create a fully rectified, sky-subtracted
2D spectrum for each slitlet, the sky being subtracted by simply
subtracting the 'nodded' spectrum from the 'un-nodded' observation. At
this point, since 1D extraction had not been implemented for N\&S in
the {\sc COSMOS} package, we used our own custom written {\sc IDL}
routines (detailed below) to extract 1D spectra.

Each slitlet was searched for a peak corresponding to the galaxy continuum
using the routine {\sc peakinfo} taken from the SDSS spec2D
package\footnote{{\tt http://spectro.princeton.edu}}, after collapsing
the image to 1D in the spectral direction.  If a peak was found then a
corresponding {\it negative} peak was searched for, at approximately
the 'nod' distance away from the positive peak.  If this approach
failed to yield two consistent peaks, then a smaller searchbox was used
in the wavelength direction, and this box shifted until a pair of peaks
were located. If a pair of peaks could not be found, then only the
largest positive peak was selected.

For each slit, each spectrum was extracted by weighting the data around
the centre of the peak by a Gaussian of the width fitted by {\sc
  peakinfo}.  Each slitlet typically contained two observations of
each object (the positive and negative spectra from the N\&S
observations) and two exposures for each mask.  In order to combine
these 1D spectra, the data were coadded after scaling by the exposure
time and rejecting highly deviant positive points (or negative in the
negative spectra) corresponding to cosmic ray hits.  This simple sigma
rejection removed a large fraction of cosmic rays, but some residual
hits were rejected later, manually, by comparing the individual
2D extractions for each slit.

\subsection{GMOS reduction}
Both sets of GMOS data were reduced using the standard Gemini {\sc iraf} routines\footnote{\tt http://www.gemini.edu/sciops/data/dataSoftware.html} to bias subtract, flatfield and wavelength calibrate the data in a manner similar to that described above.  The i{\sc GDDS} package \citep{2004AJ....127.2455A} was used to interactively trace the 2D spectra and extract 1D spectra.

\section{Analysis}

\subsection{Redshift determination}

Redshifts were determined using the {\sc rvsao} package
\citep{1998PASP..110..934K} within {\sc iraf}. Firstly all 1D spectra were
cross-correlated (using {\sc xcsao}) with a range of spectral templates
including the E/S0, Scd and Sab galaxies used by the CNOC
collaboration \citep{Yee:1996lk}, and the SDSS composite quasar spectrum
\citep{2001AJ....122..549V}.  Next, emission lines were searched for
with {\sc emsao} using the cross-correlation redshift as an initial
estimate of the redshift.  This task was run interactively and the
redshift adjusted manually, in cases where the automated redshift was
clearly incorrect, by fitting to emission or absorption
features. The 1D and 2D spectra were simultaneously inspected, in order to
confirm the reality of faint features.  In the case of the IMACS (nod \& shuffle) data,
the 2D spectra of the combined and the individual exposures were
'blinked' in order to check for residual cosmic rays masquerading as
emission lines.  These were easily rejected by noting the presence of a
bright feature in one exposure only.  In addition, a number of emission
line only spectra which were not correctly identified and extracted
(since no continuum peak was found) were found with this interactive
process. The 1D spectra were displayed with features overplotted and
visually inspected and then a quality flag assigned
(Table 2).  Examples of randomly selected spectra from
each of the quality classes are shown in Fig.~1.

For the GMOS nod \& shuffle data, redshifts were estimated in i{\sc GDDS} by overplotting a variety of templates on the 1D spectra at various trial redshifts.  This technique was also compared with the {\sc rvsao} method and found to give consistent results.  The benefit of i{\sc GDDS} is the ease with which a variety of different templates can be tested whilst simultaneously examining the 1D and 2D spectra to confirm the reality of low signal-to-noise features.

\begin{deluxetable}{lp{5cm}c}
\label{table:flags}
\tablecaption{Summary of redshift quality flags.}
\tablewidth{0pt}
\tablehead{
\colhead{Flag} & \colhead{Comments}
}
\startdata
1    & Secure redshift &     \\
2    & Probable redshift (e.g., one emission line with probable support
or several weak features) &    \\
3    & One emission line only, but no support (assumed [O{\sc
  ii}]$\lambda$3727) &    \\
4    & Possible redshift, but unconvincing & \\
5    & No redshift &     \\
\enddata
\end{deluxetable}

\begin{figure}
\epsscale{.80}
\plotone{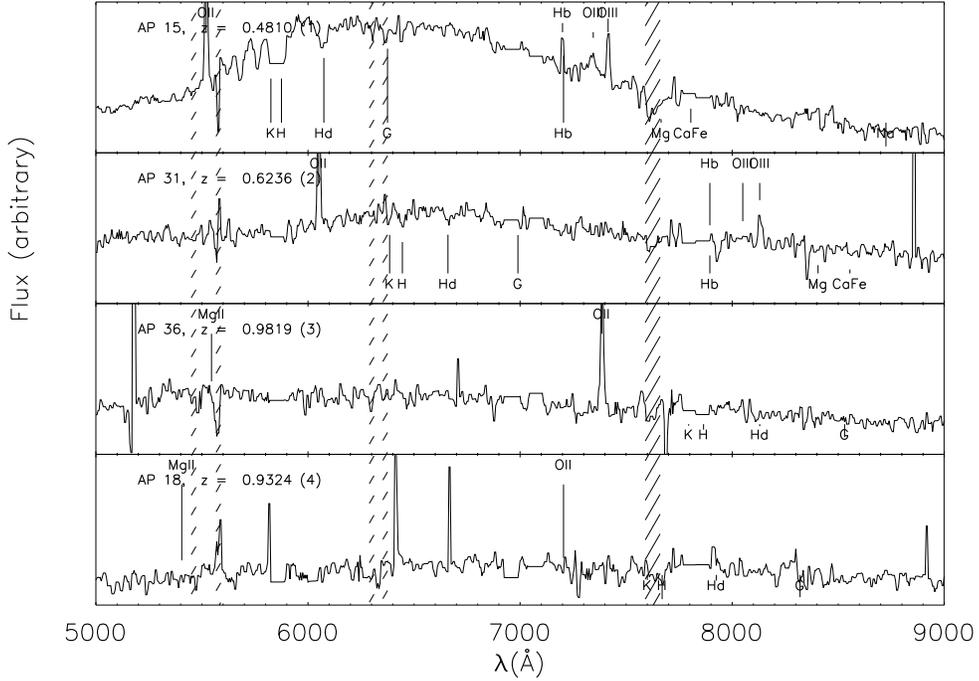}
\caption{Example LDSS-2 spectra from each quality class. The redshift and
  quality flag is indicated in the upper left of each panel. From upper to lower
  panel, spectra of redshift quality flag 1,2,3,4 (Table 2)
  respectively. Fluxes are in arbitrary units and spectra have not been flux calibrated; labelled lines indicate the expected positions of common emission and absorption features;
  hatched areas denote regions potentially contaminated by strong night sky
  line residuals.}
\end{figure}

\begin{deluxetable}{lccccccp{8cm}}
\tabletypesize{\scriptsize}
\rotate
\tablecaption{ Cluster properties and spectroscopic data.  Columns show: cluster name, detection significance in catalogue, \bgc~(optical richness), RA and Dec of mask centre, photometric redshift, spectroscopic redshift and details of galaxies identified with possible overdensities in redshift space. For overdensities of more than 10 galaxies, velocity dispersions have been calculated in \S4.3 and the number of members calculated after 3-$\sigma$ clipping are listed.}
\label{tab:summ}
\tablewidth{0pt}
\tablehead{
\colhead{ID} & \colhead{$\sigma_{\rm RCS}$} & \colhead{\bgc}\tablenotemark{a} & \colhead{$\alpha$ (J2000)} & \colhead{$\delta$ (J2000)} & \colhead{z$_{phot}$} & \colhead{z$_{spec}$} & \colhead{Comments}
}
\startdata
\cutinhead{LDSS-2}
RCS033414-2824.6 & 4.1 & 1270$\pm$305 & 03:34:12.3 & -28:24:16 & 0.683 & 0.668 & 20 class 1 members, 26 class 1-3 members give $\sigma$=$(300\pm60)$ \kms~\\
RCS034850-1017.6 & 3.2 & 710$\pm$330 & 03:48:49.7 & -10:17:45 & 0.879  & -----\tablenotemark{b} & ---- \\
RCS043938-2904.7 & 4.7 & 1590$\pm$460 & 04:39:38.0 & -29:04:55 & 0.937 & 0.869 & 3 class 1\&2 redshifts within 1000 \kms\tablenotemark{c}\\ 
               & &   &         &             &      & 0.974 & 4 class 1-3 redshifts within 1500 \kms\\
RCS044111-2858.2 & 3.3  & 830$\pm$470 & 04:41:11.4 & -28:58:15 & 1.079 & 0.950 & 3 class 1\&2 redshifts within 400 \kms\\
RCS110246-0426.9 & 4.0  & 930$\pm$250 & 11:02:45.9 & -04:26:53 & 0.737 & 0.723 & 5 class 1 redshifts within 1400 \kms\\
RCS110634-0408.9 & 4.0  & 660$\pm$230 & 11:06:33.3 & -04:09:03 & 0.805 & 0.823 & 5 class 1 redshifts within 600 \kms\\
RCS110708-0355.3 & 3.2 & 300$\pm$190 & 11:07:17.9& -03:55:04 & 0.918  & 0.825  & 5 class 1  redshifts within 800 \kms~(note: this cluster does not appear in the December 2005 cluster catalogues and the values are taken from the December 2004 catalogue)\\
RCS110723-0523.3 & 3.5 & 980$\pm$300 & 11:07:22.8 & -05:23:49 & 0.767 & 0.735 & 20 class 1 members, 23 class 1-3 members give $\sigma$=$(700\pm300)$ \kms\\
\cutinhead{IMACS}
RCS035231-1020.7 & 4.0  & 710$\pm$230 & 03:52:31.0 & -10:20:42 & 0.816  & 0.709 & 12 class 1-3 members give $\sigma$=$(600\pm370)$ \kms\\
RCS043938-2904.7 & 4.7 & 1590$\pm$460 & 04:39:38.0 & -29:04:55 & 0.937  & 0.960 & 10 class 1-3 redshifts within 2400 \kms~\tablenotemark{a}\\
\cutinhead{GMOS-N}
RCS141658+5305.2 & 4.7  & 3110$\pm$800 & 14:16:58.7 & +53:05:15  & 1.150 & 0.9682 & 9 class 1 redshifts within 700 km/s [tentative $\sigma$=$(1030\pm1000)$ \kms]\\
                                 &      &        &                     &                      &            & 0.8945 & 6 class 1 redshifts within 300 km/s  [tentative $\sigma$=$(240\pm70)$ \kms)]                        \\
RCS162009+2929.4 & 4.4 & 930$\pm$240 & 16:20:09.4 & +29:29:26 & 0.885 & 0.8696 & 12 class 1 members, 13 class 1-3 redshifts give $\sigma$=$(1050\pm340)$ \kms\\
\enddata
\tablenotetext{a}{\bgc~is the value which was measured directly from the survey data.  No {\it a posteriori} correction has been applied here for overlapping red-sequences in the cases of  RCS043938-2904.7 and RCS141658+5305.2 (see \S4.1.1 and \S4.1.4).
}
\tablenotetext{b}{The spectroscopy for RCS034850-1017.6 appears to have been insufficiently deep to confirm a cluster at this redshift (see \S4.1.3).
}
\tablenotetext{c}{For RCS043938-2904.7 we can add data from \cite{2004ApJ...617L..17B}.  For the composite LDSS-2, IMACS, FORS2 dataset we identify a system with a mean redshift of 0.955 with  
20 redshifts within 2400 \kms, but see discussion in \S4.}

\end{deluxetable}

\subsection{Cluster confirmation}

The simplest and
most conservative test to confirm cluster candidates is to plot
redshift histograms for the secure (class 1) redshifts and look for
overdensities.  Figs.~2 and 3 show the large scale redshift histograms for
each cluster field, shaded according to the redshift quality.  A fixed
binsize of z=0.01 is used, which translates to a width in rest-frame
velocity of $\sim$1700 \kms~to $\sim$1500 \kms~at redshift 0.6 and 1.0
respectively, the approximate range for the clusters considered here.
This velocity difference corresponds to approximately the velocity
dispersion ($\sigma$) of a rich cluster of galaxies covering the 1-2-$\sigma$ range; or
a poorer cluster or rich group over 2-3-$\sigma$.

This immediately yields 7 fields containing at least one peak comprising 5 or more secure
redshifts: RCS033414-2824.6, RCS110634-0408.9, RCS110708-0355.3, 
RCS110723-0523.2, RCS035231-1020.7, RCS141658+5305.2 and RCS162009+2929.4.  The velocity distribution in the vicinity of the
overdensity is plotted in the inset panels.  This time a fixed rest-frame velocity bin size of 200\kms~is used.

A less conservative test is used on the remaining clusters. We
include spectra of other quality flags when searching for
overdensities.  We find the maximum (i.e., poorest) quality spectra
which needed to be included to produce at least 3 galaxies within a
bin.  The resulting maximum velocity differences and quality
flags for these systems are presented in Table 3.  All the
systems except for RCS034850-1017.6 yield at least 3 galaxies with
class 1-3 redshifts within 1300 \kms.  We discuss the significance of
such overdensities in \S\ref{sec:discussion}.

For the two clusters comprising $\gsim$20 members, it is
reasonable to calculate a velocity dispersion.  We use the biweight
scale estimator as recommended by \citet{bfg90} for n$\sim$10-20
galaxies, and a jackknife estimate of the uncertainty.  Following the procedure of \cite{Danese:1980ng}, we subtract in quadrature 100 \kms, representing the typical uncertainty in an individual redshift measurement\footnote{This value is based on our experience with similar datasets, since we do not have enough repeat measurements within this work to determine the measurement uncertainties internally, but see \S4.1.1}.  For RCS033414-2824.6 we find a rest-frame velocity dispersion of
$\sigma=(300\pm60)$ \kms~and for RCS110723-0523.2, $\sigma=(700\pm300)$
\kms, using only class 1 redshifts.  Including class 1-3 redshifts
the values are: $\sigma=(400\pm100)$ \kms~and $\sigma=(600\pm150)$
\kms, respectively.  

Two further clusters contain $\gsim$10 members and so it is worth attempting to estimate velocity dispersions for these systems too, although the uncertainties will be higher.  The cluster RCS162009+2929.4 has 13 class 1-3 spectroscopic members.  These yield $\sigma=(1100\pm 350)$ \kms. 
  In addition, RCS035231-1020.7 has 11 class 1 spectroscopic members, giving a velocity dispersion of $\sigma=(600\pm300)$ \kms.  

We note that when dealing with $\sim$20 cluster redshifts, a possible source of systematic uncertainty may be structure nearby in redshift space, unresolved along the line of sight.  For example, \citet{Gal:2004gt} found that, in a supercluster at z$\sim$0.9, previous velocity dispersions measurements based on $\sim$20 members overestimated the velocity dispersion by $\sim$30-40\%, as compared with their factor of two larger spectroscopic dataset.  This was due to galaxies in nearby foreground and background groups being incorporated into the estimate for the velocity dispersion of the cluster.  This 30\%-40\% uncertainty should probably represent an upper limit to the systematic error, since the probability of incorporating additional structure is higher in such a supercluster environment.  We also note that our quoted uncertainties are already of order this amount.

\begin{figure*}
\epsscale{1.02}
\plottwo{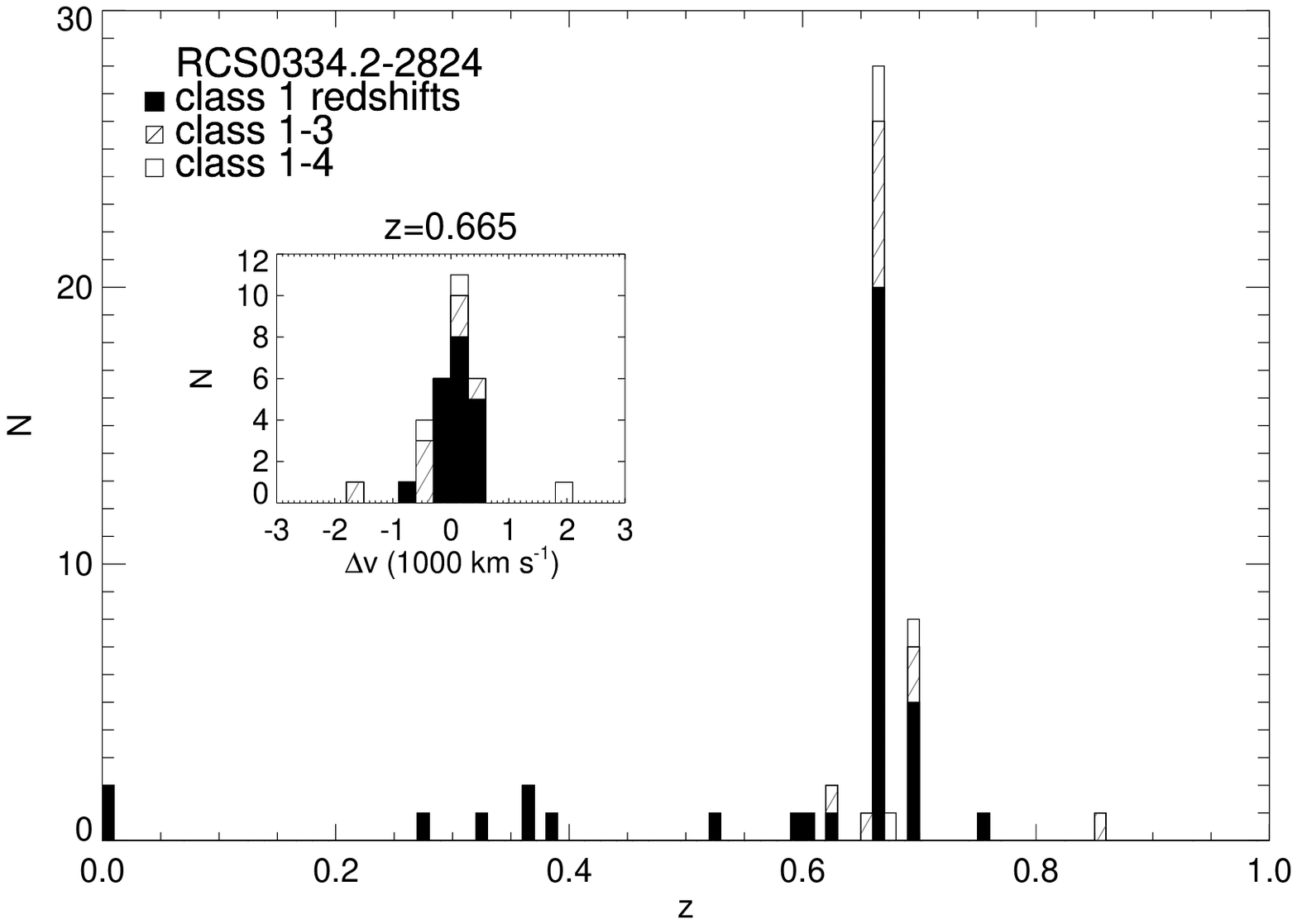}{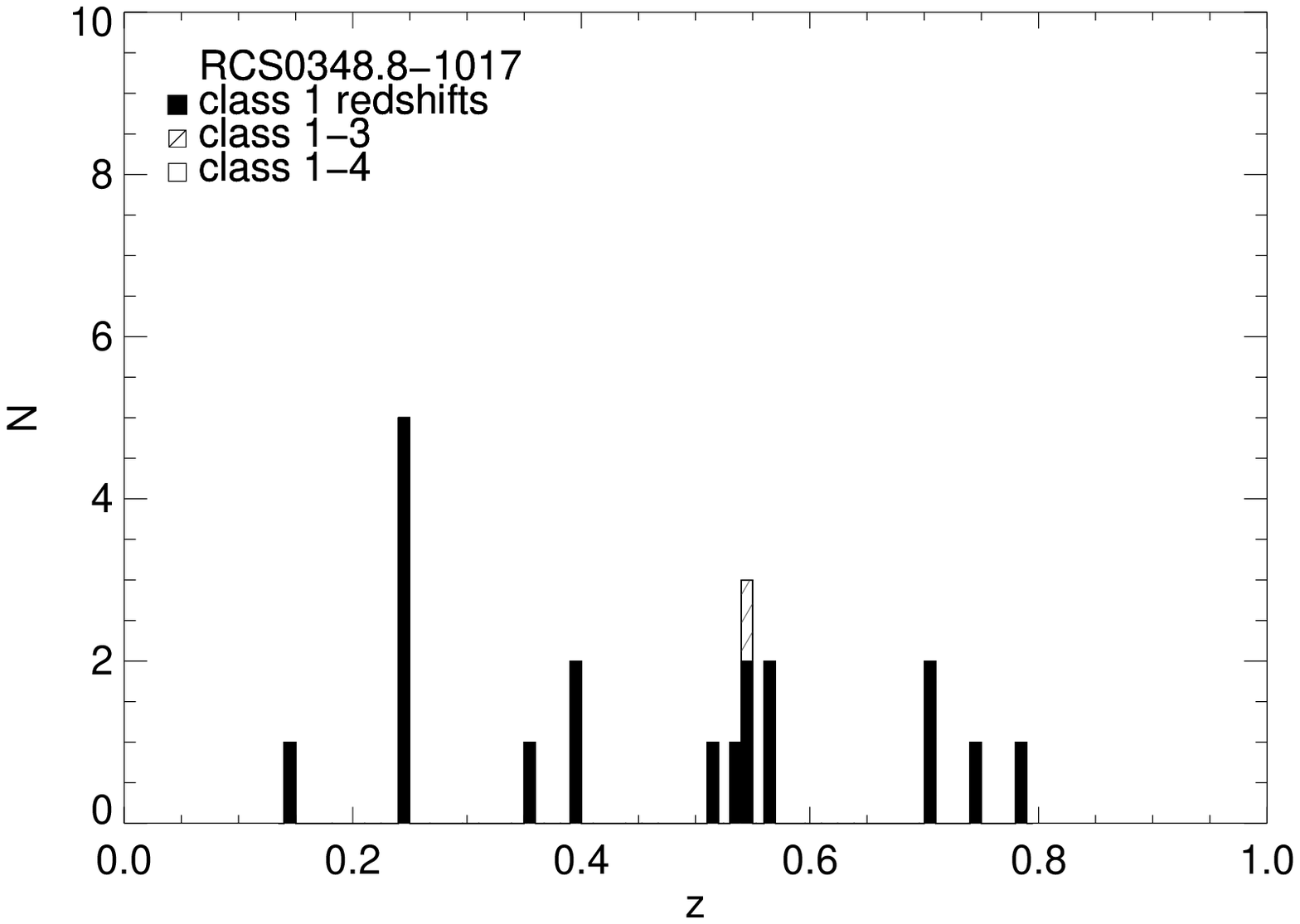}
\plottwo{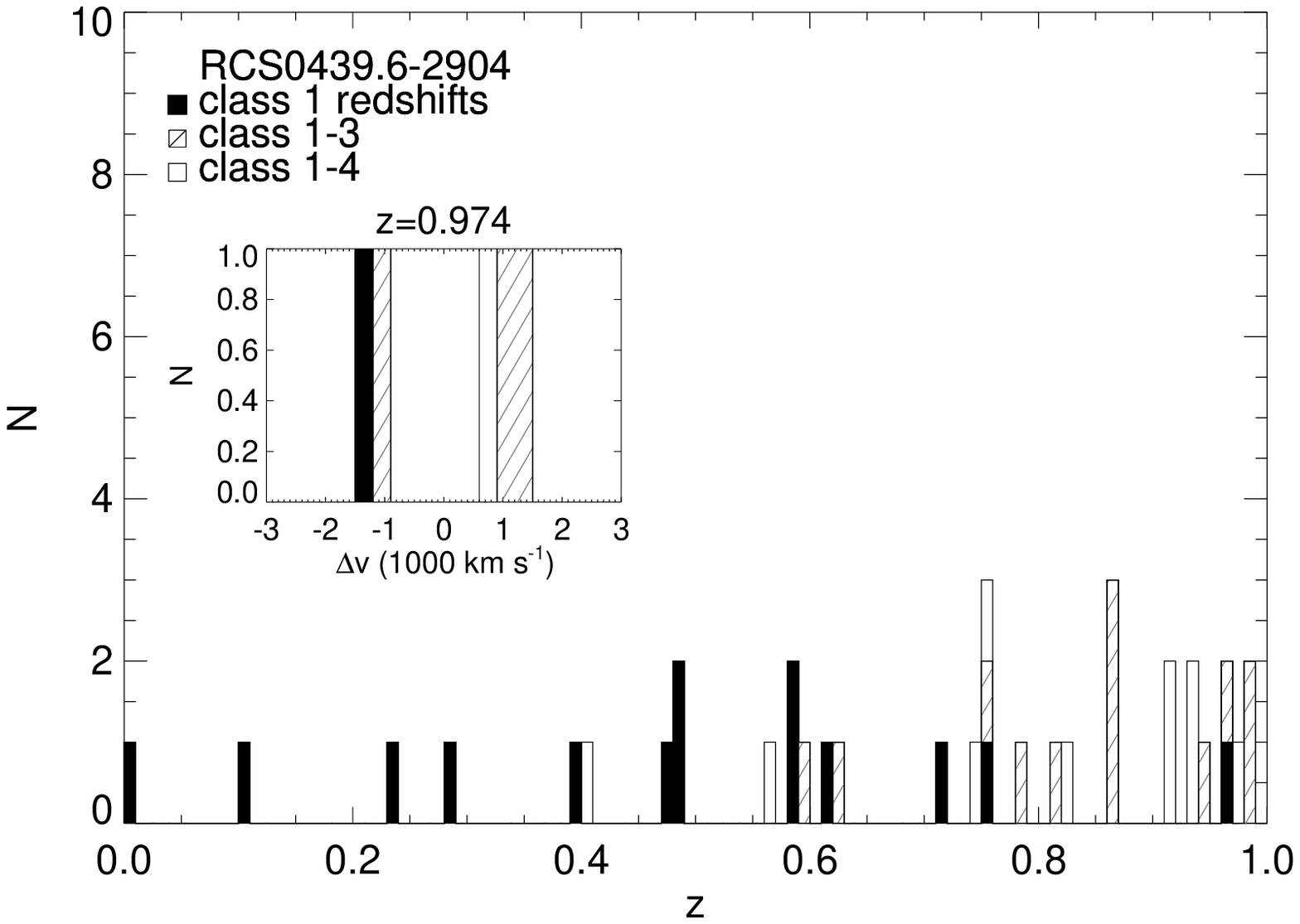}{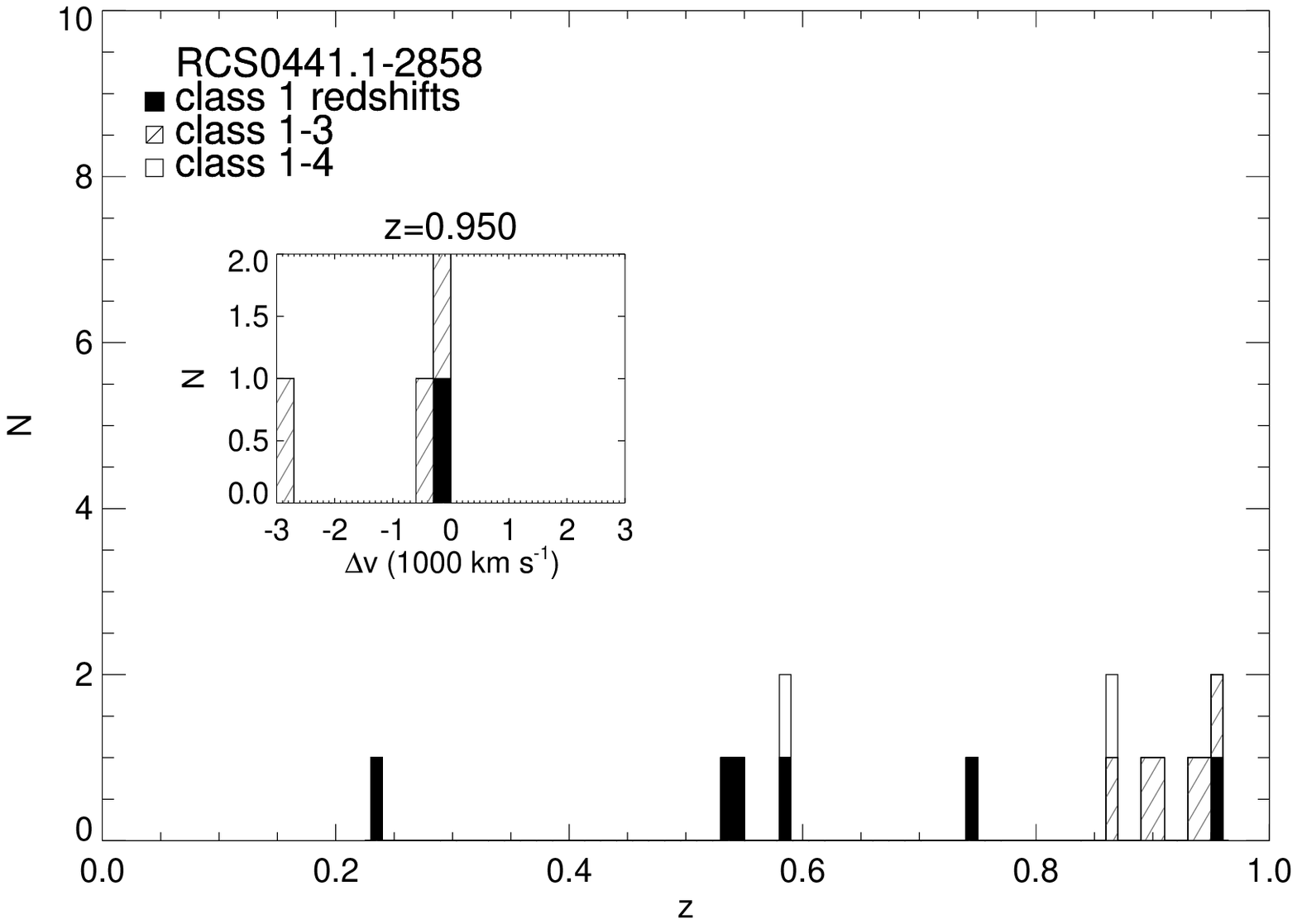} 
\plottwo{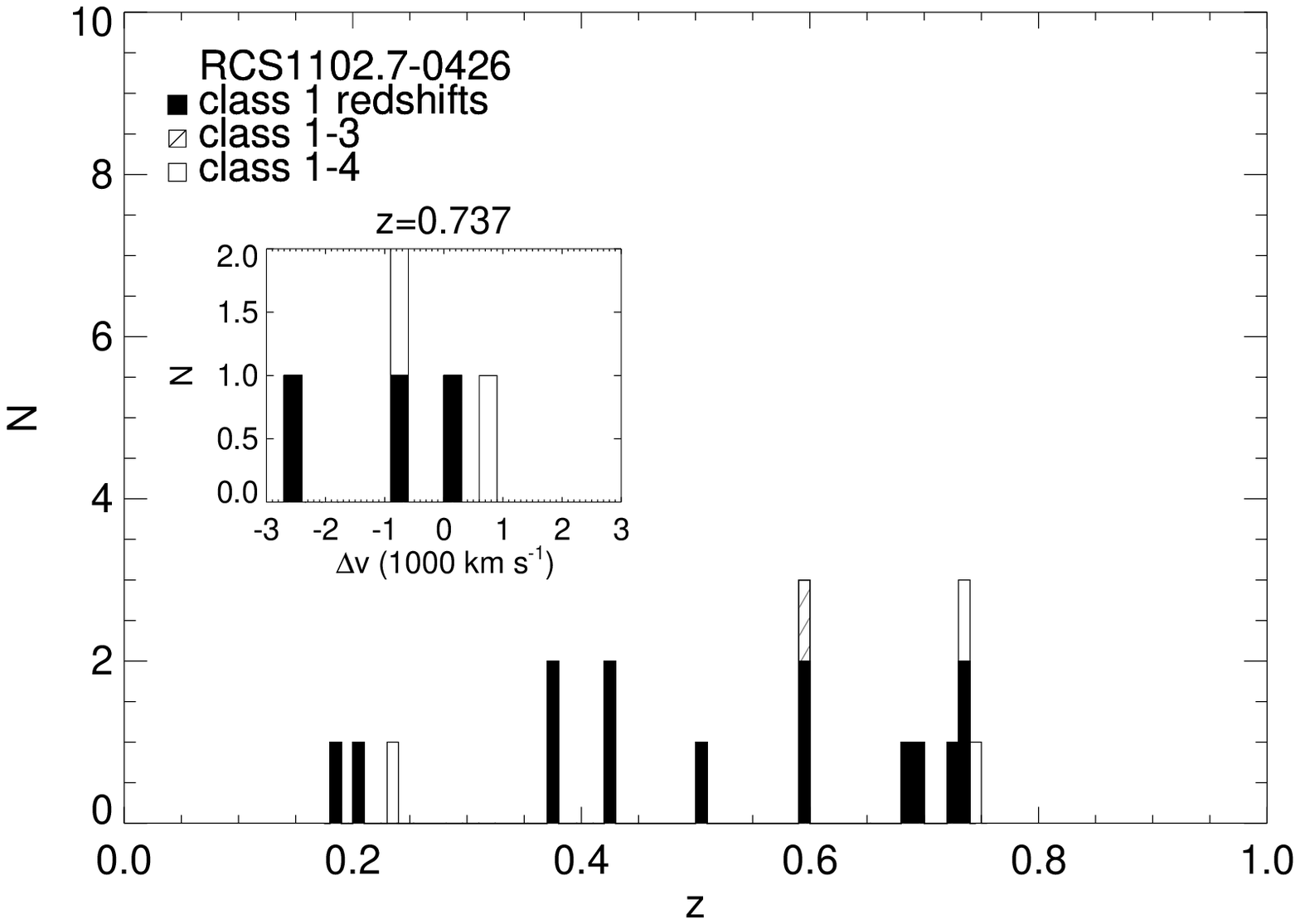}{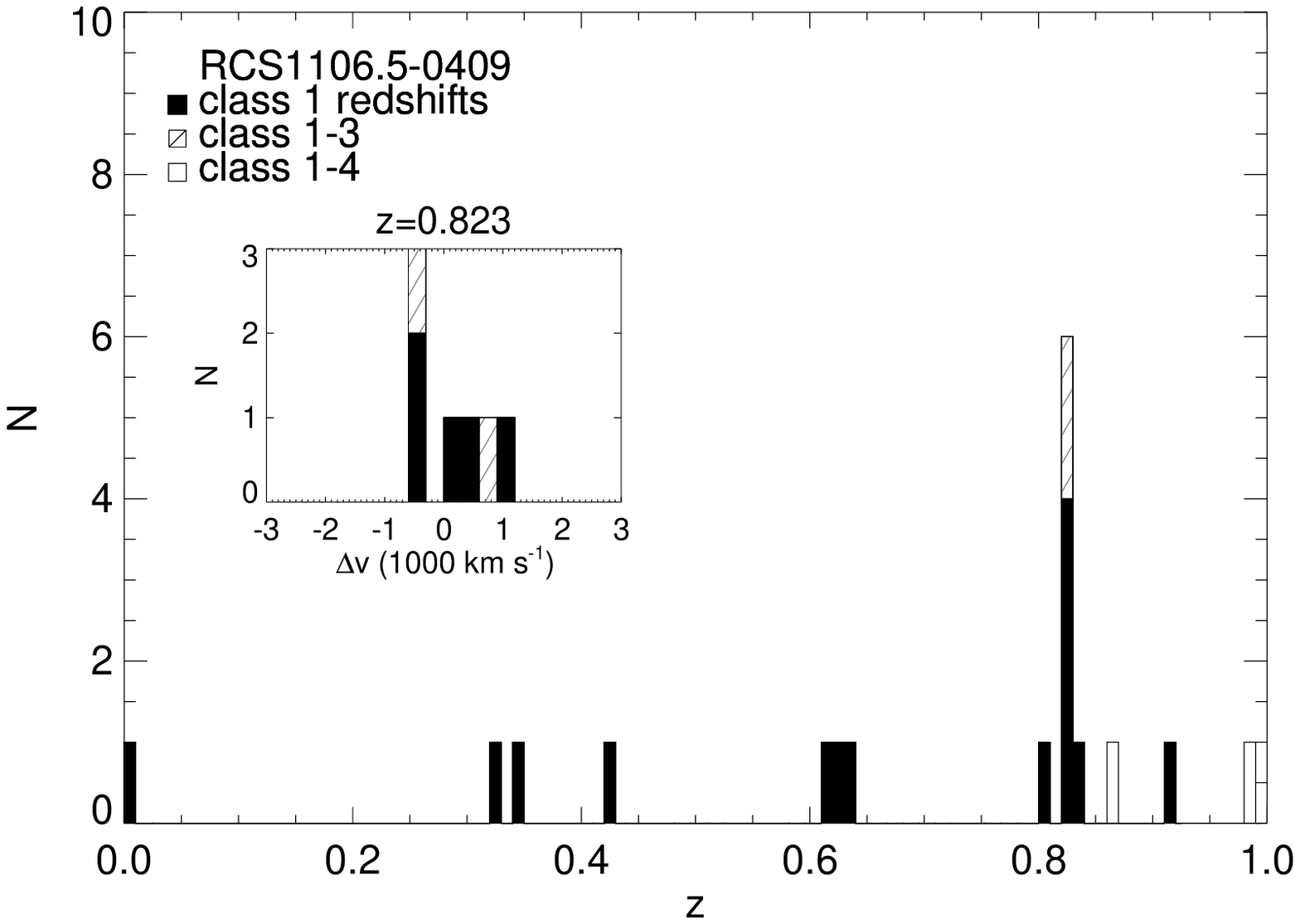}
\plottwo{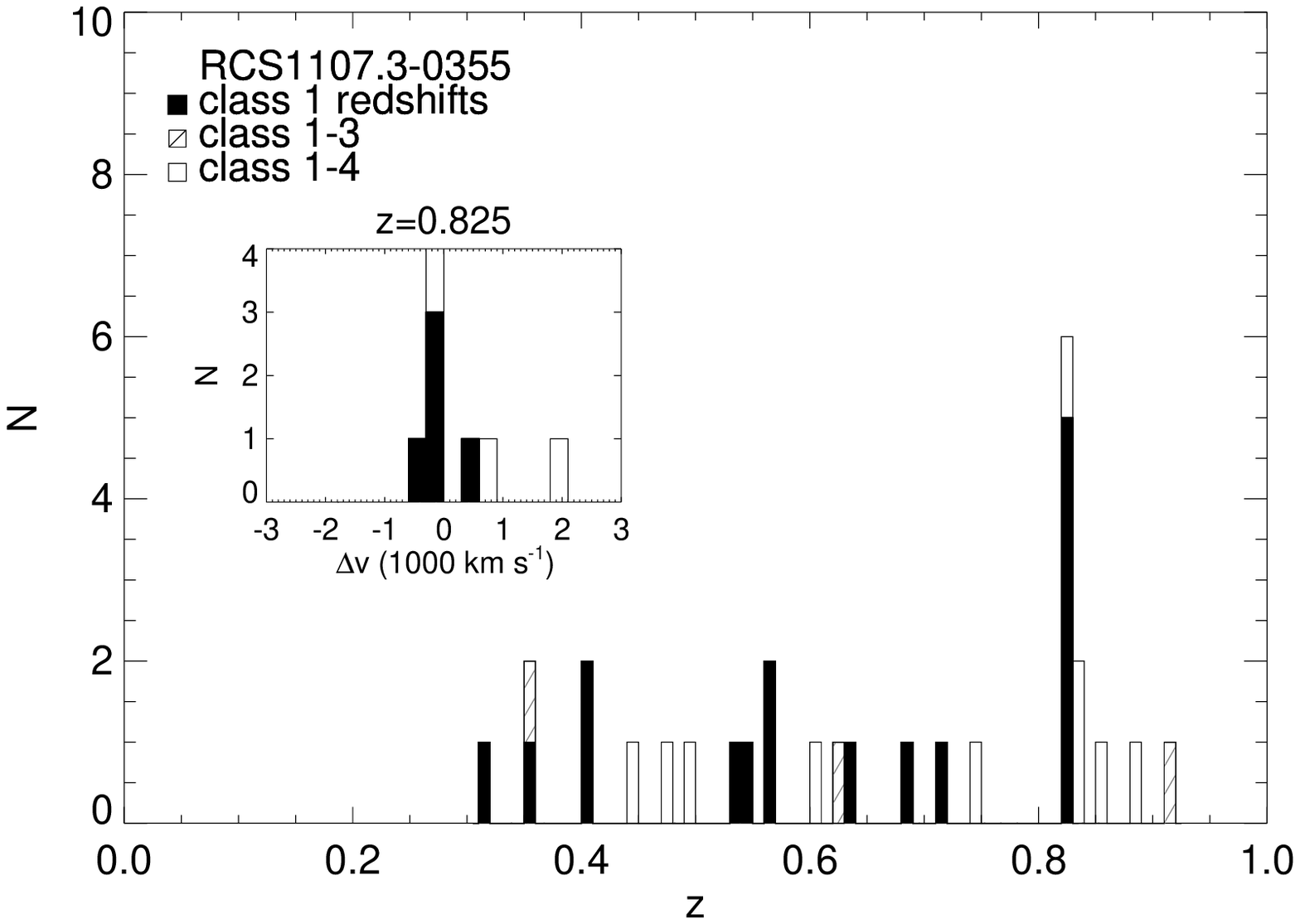}{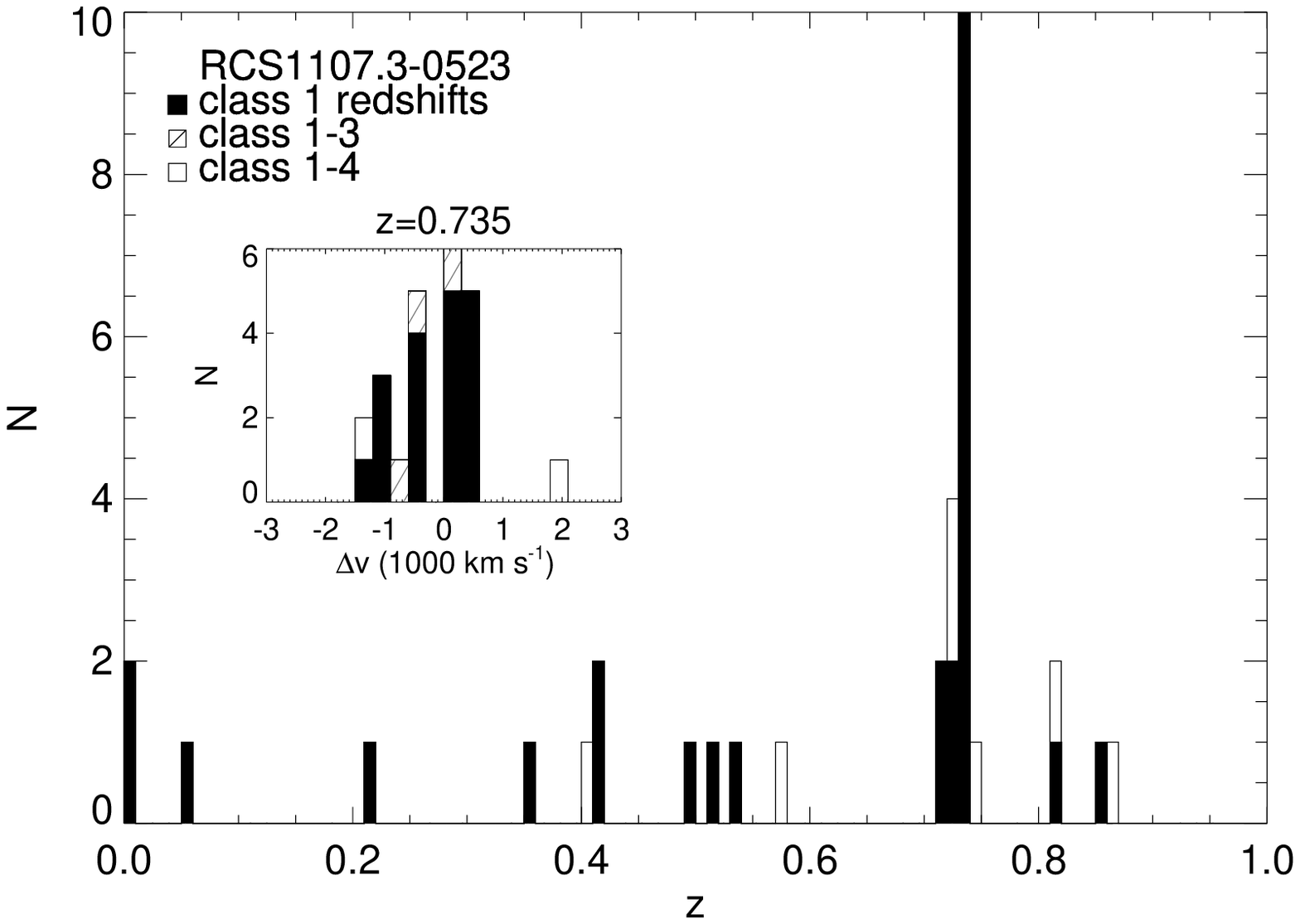} 
\caption{\footnotesize
Redshift histograms for the RCS clusters observed with LDSS-2.  Spectra are labelled according to their redshift quality flag as described in Table 2.  Bins are 0.01 in redshift, corresponding to $\sim$1700\kms~at z$=$0.6 to $\sim$1500\kms~at z$=$1.0.  Insets show expanded views in rest-frame velocities around overdensities in redshift space.  }
\end{figure*}

\begin{figure*}
\epsscale{1.1}
\plottwo{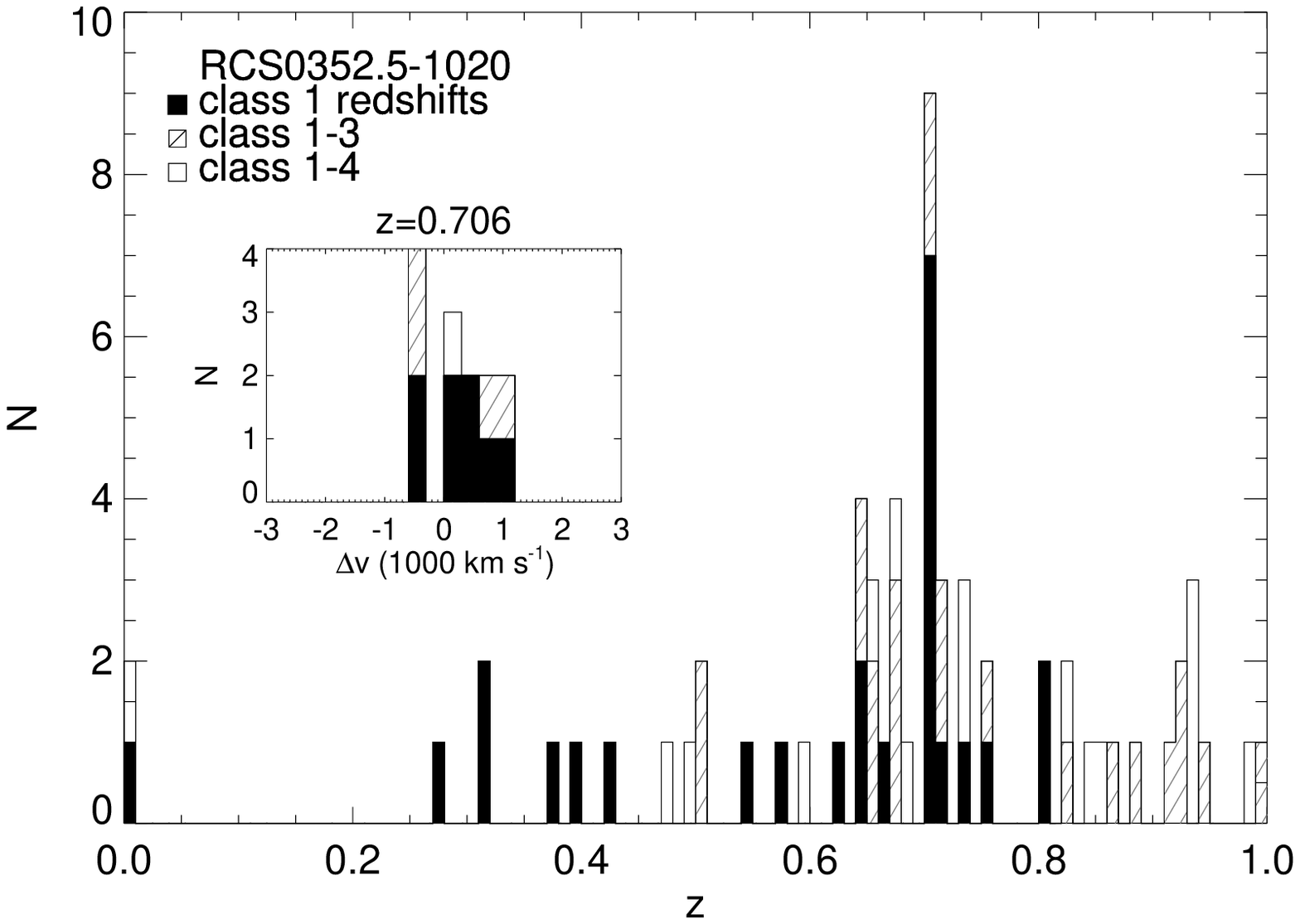}{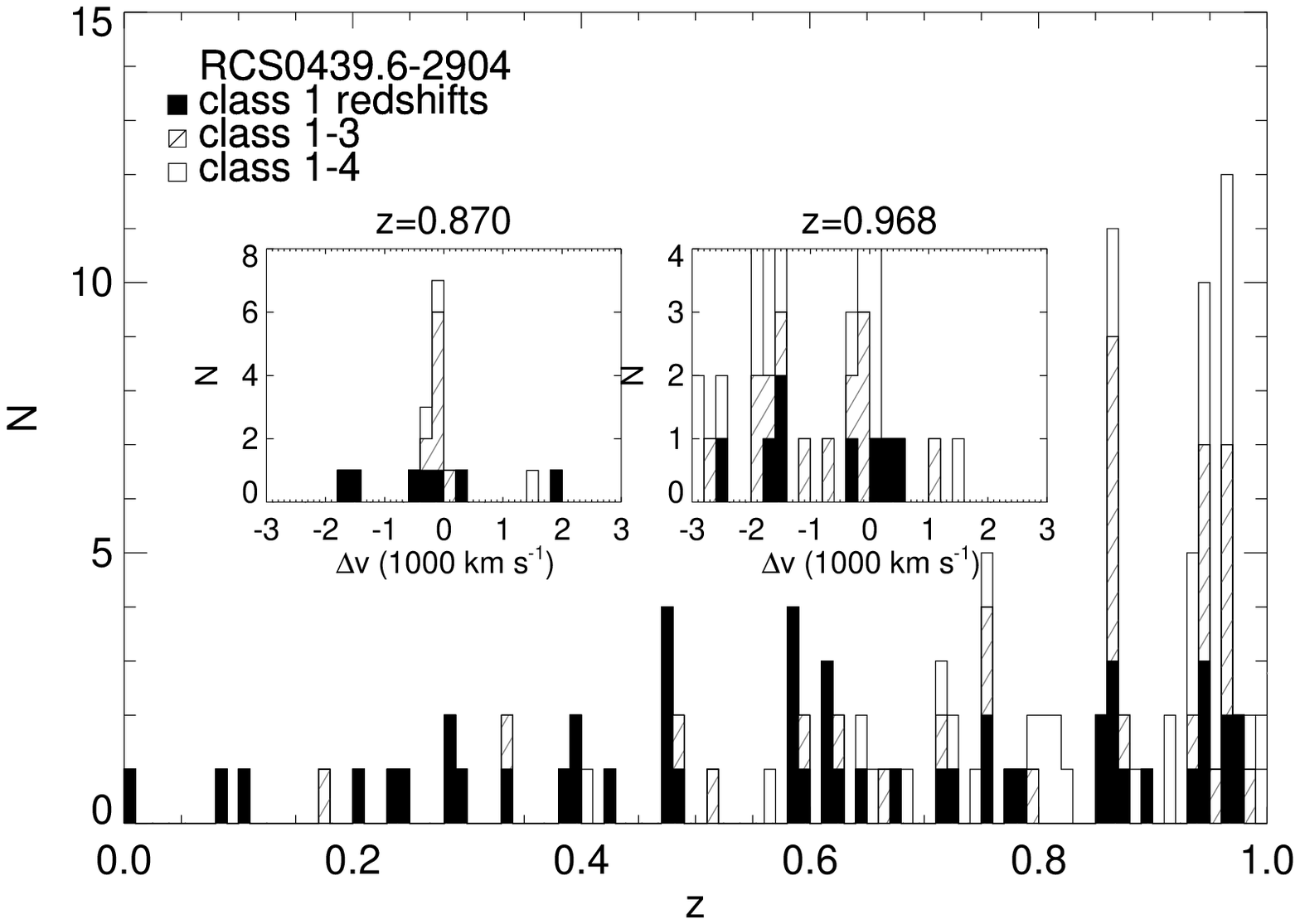}
\plottwo{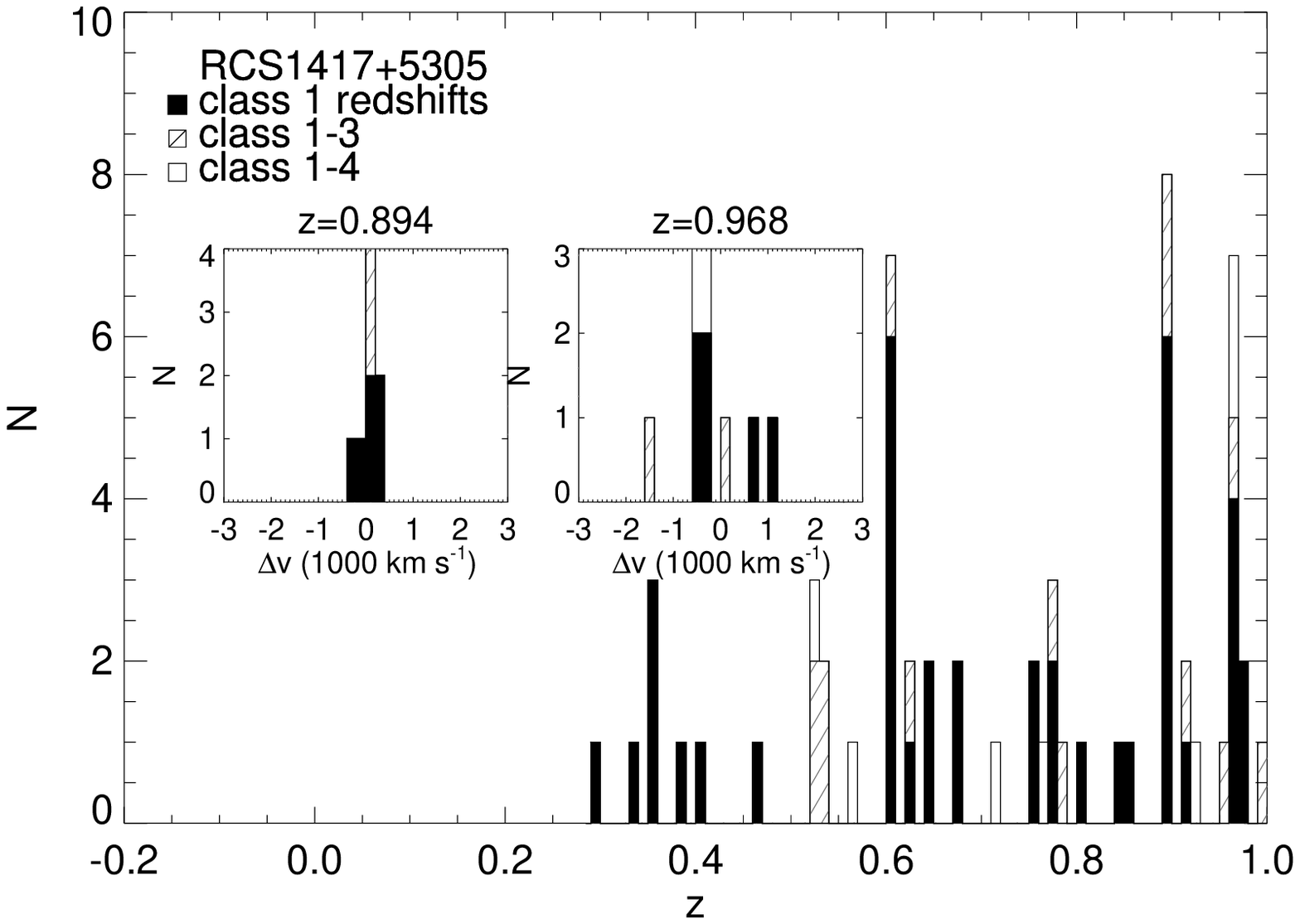}{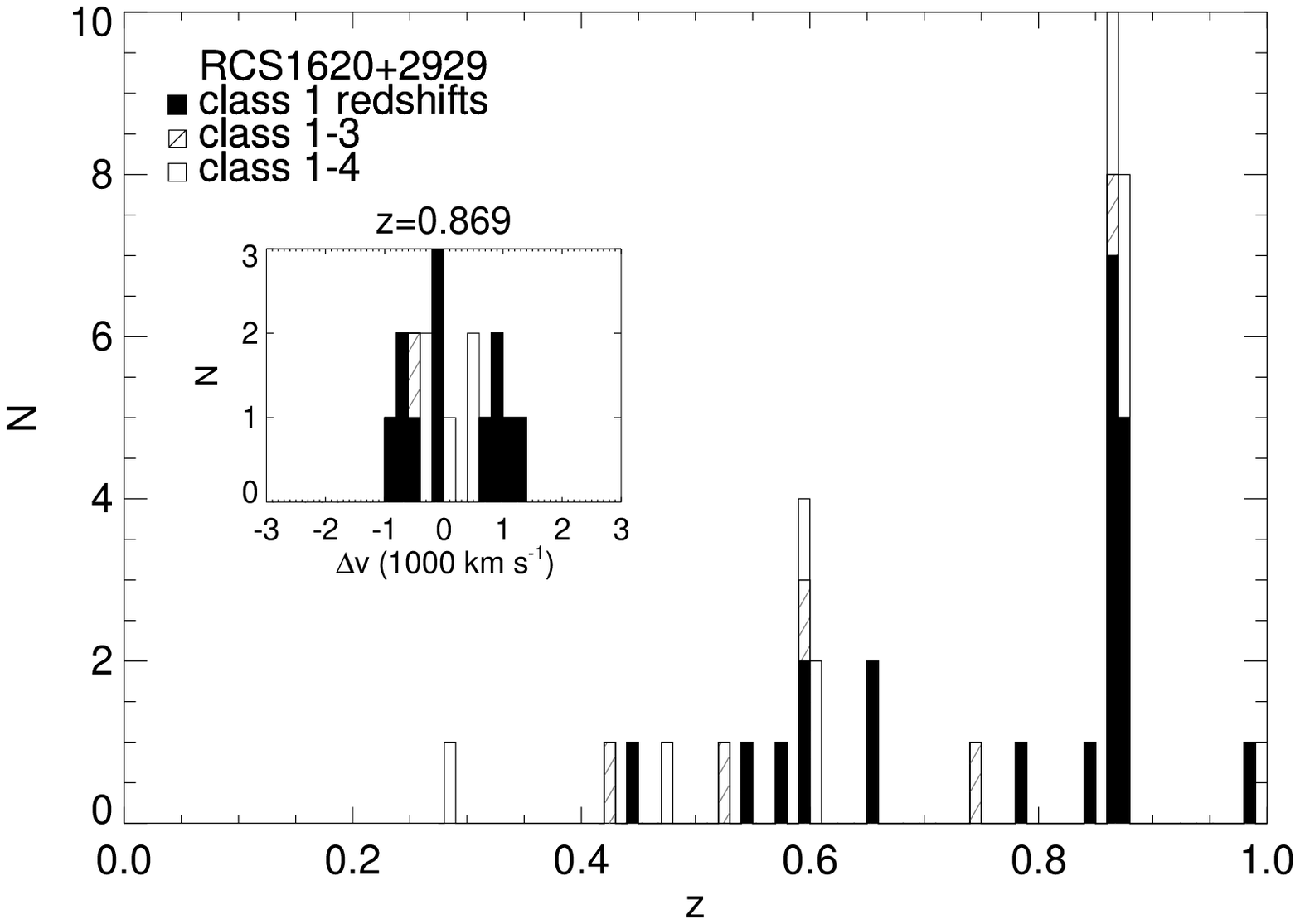} 
\caption{As for Fig.~2, but for clusters observed with IMACS (upper panels) and GMOS-N (lower panels).
The plot for RCS0439.9-2904 (upper right) includes all available data from LDSS-2, IMACS and FORS2.}
\end{figure*}

\section{Results \& Discussion}
\label{sec:discussion}

\subsection{Individual clusters}

\subsubsection{RCS043938-2904.8}

The field of RCS043938-2904.8 offers the possibility of testing the
accuracy of redshift measurements of
duplicate objects taken with the different instruments as it was observed with both LDSS-2 and IMACS.  
Furthermore, additional data are available, taken with FORS2 on the VLT \citep{2004ApJ...617L..17B}.
Three objects from the IMACS mask were also observed with FORS2.
In one of these the object lies on the slit edge in the FORS2 data and cannot be reliably
extracted; the next object has a redshift flag of 4 in both datasets,
but still yields a pleasingly consistent redshift (1.121 from IMACS; 1.119
from FORS2) within 300 \kms~rest-frame; and the third is an emission line
galaxy at z=0.2945, agreeing to better than 30 \kms~between the two
instruments.  The fact that a class 4 redshift appears to have been
reproduced, albeit with a larger uncertainty than the secure
measurements, suggests that the classification system is reliable, if
somewhat cautious.

This field also allows us to test the reproducibility of
structures identified with the different instruments.
Although no obvious large overdensity is
seen in the LDSS-2 data, four galaxies (redshift classes 1-3) are seen
within 1300 \kms~of each other at z=0.960 (inset of Fig.~2).  A second
possible peak of three galaxies at z=0.869 is also seen.  Prominent peaks are
visible in the redshift histograms near both these positions in the
IMACS data (Fig.~3).  This reinforces the idea that marginal confirmations of
overdensities comprising only three or four galaxies will be confirmed
with supplemental spectroscopy.

Close inspection of the redshift histogram in Fig.~3 (right inset panel) reveals that the overdensity at z$\sim$0.96 actually appears to comprise two peaks: one at z=0.945 and one at z=0.968.   This corresponds to a rest-frame velocity difference of 3500\kms.  Cluster mergers can reach relative velocities of $\sim$3000 \kms~\citep{Sarazin:2002rv}.  Thus it is possible that these two systems may be physically related.  This nature of this system will be discussed further in conjunction with X-ray observations in Cain et al.\ (in prep).  For now, we note that this cluster is potentially binary or comprises the projection of two clusters, and associate this system with the target of our spectroscopic observation.
  
\begin{figure*}
\epsscale{1.3}
\plottwo{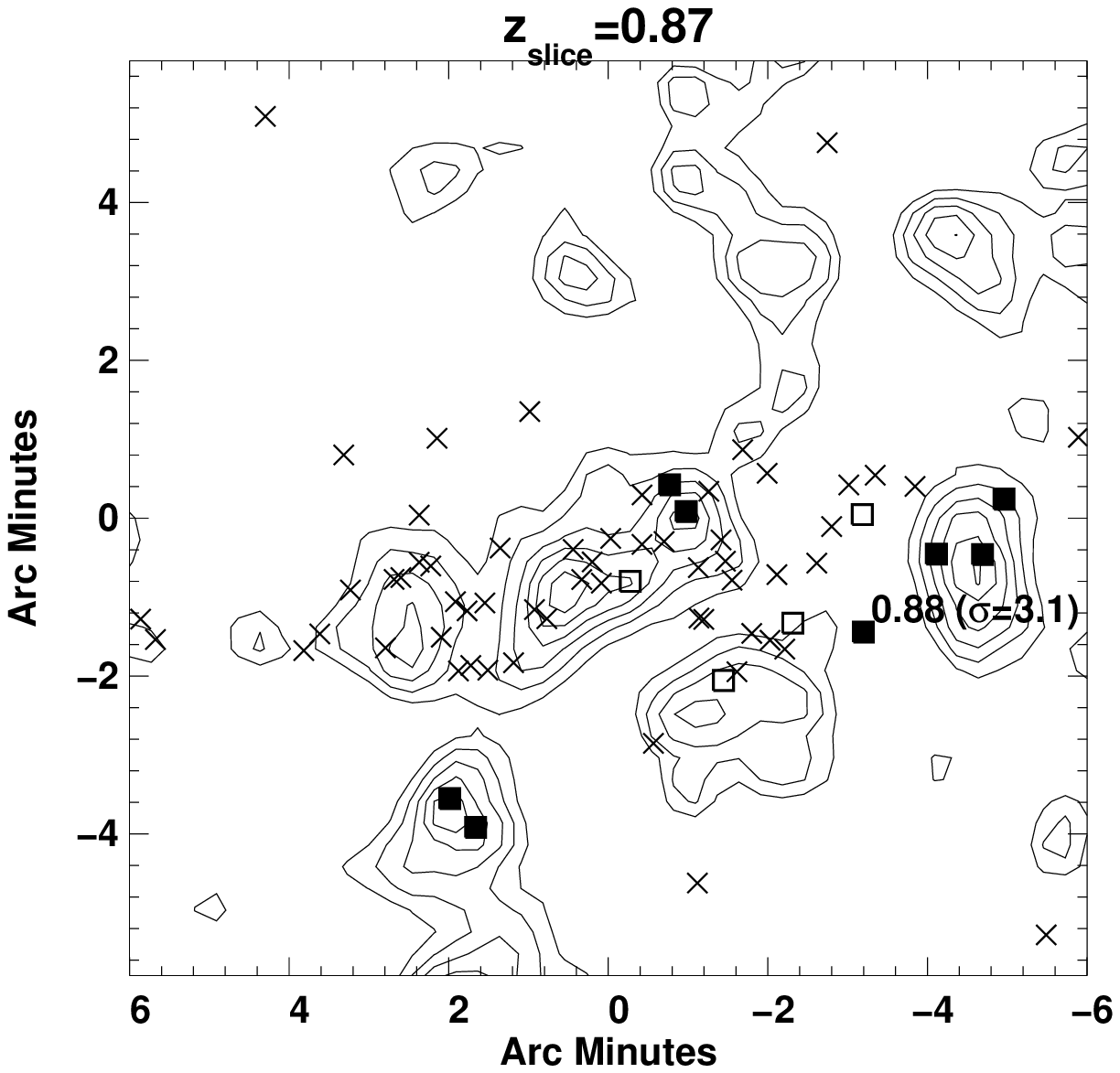}{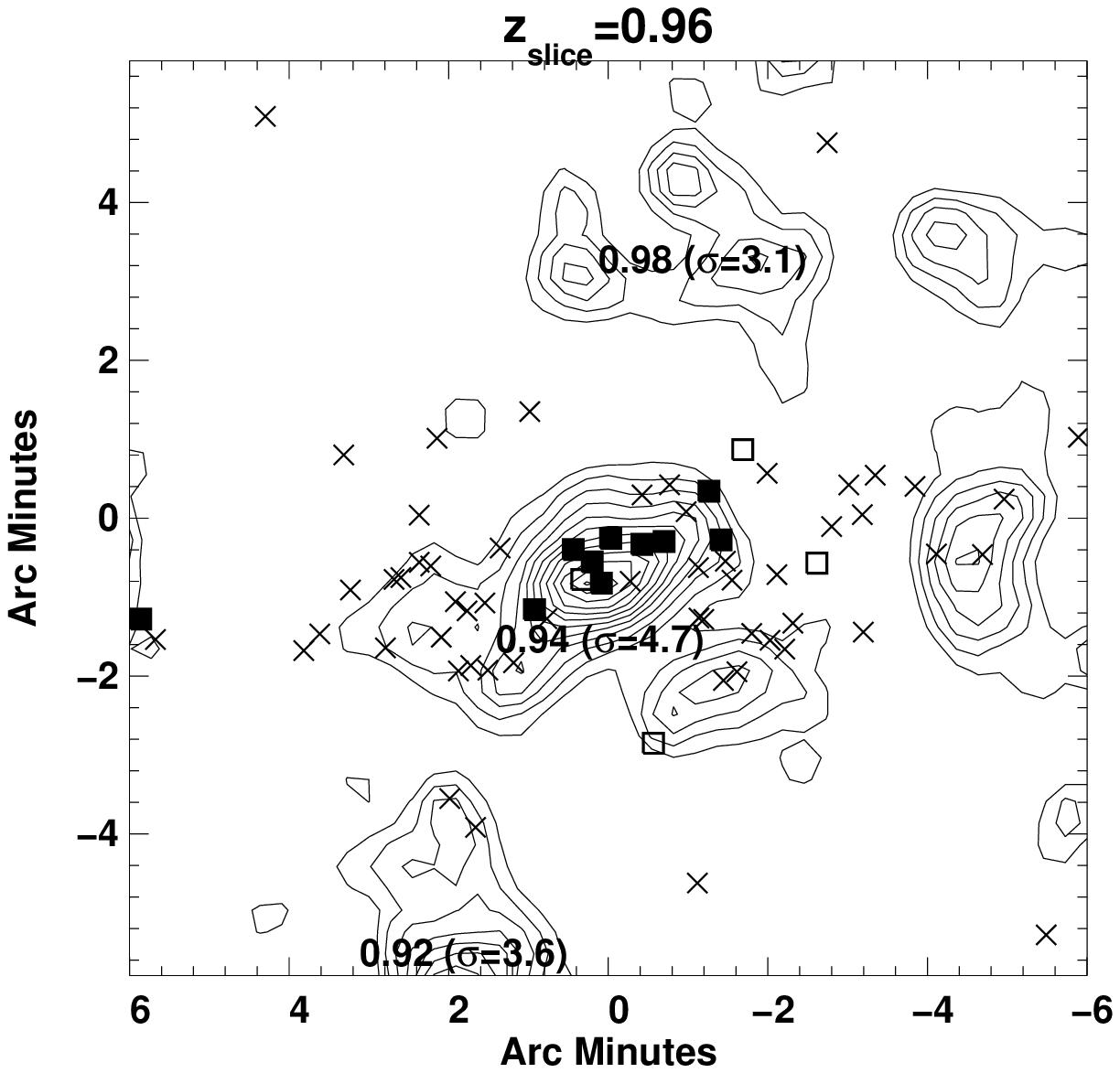}
\caption{Significance maps from the RCS technique for RCS043938-2904.8 with spectroscopic members overlaid.  Contours show the significance of structures identified in the RCS technique.  Contours are in intervals of 0.3-$\sigma$, starting at 1.5-$\sigma$.  Labels on the highest peaks identify the redshift and significance of peaks identified as cluster candidates.  The left panel is for z$_{phot}$=0.87 and the right is for z$_{phot}$=0.96 corresponding approximately to the two peaks identified in the histogram of Fig.~2.  Crosses denote spectroscopic non-members of each structure, filled squares show red members with redshifts compatible with the redshift "slice" and open squares denote blue spectroscopic members.  There is clearly a large overdensity of red galaxies associated with the cluster candidate at 0.94.  Galaxies in the z$_{spec}=$0.87 slice are not so spatially concentrated, nor are they predominantly red.  There is a possible hint of association with the z$_{RCS}=$0.88 cluster candidate just outside the area covered for spectroscopy.  See text for further discussion.}
\end{figure*}

The red-sequence cluster-finding technique offers the possibility to disentangle multiple structures along the line of sight.  The two components of RCS043938-2904.8 (at z=0.945 and z=0.968) are too close to be separated by colour information alone, but the other peak in the redshift histogram (Fig.~3) at z$=$0.869 is potentially separable.  The $(R-z^\prime)$ colour difference between z$=$0.97 and z$=0.87$ is expected to be 0.15 magnitudes.  We note that this is larger than the intrinsic scatter of the red-sequence typically measured at these redshifts ($\sim$0.07 from {\it HST} imaging, e.g., \cite{Blakeslee:2006bd}, which becomes $\sim$0.1 magnitudes with ground-based photometric errors).  In order to study the 3D distribution of cluster candidates in this field, we examine red-sequence significance maps centred on the spectroscopic redshifts of the two main peaks (i.e., z$_{spec}\sim$0.87 and z$_{spec}\sim$0.97).  Fig.~4 displays the spatial distribution of galaxies with spectroscopic redshifts over contour maps generated from the RCS cluster detection technique.  These contour maps show the significance of overdensities of galaxies having colours and magnitudes compatible with red-sequence cluster members at the redshift of interest (see \citealt{gy00} for details). We denote such redshifts as z$_{phot}$ to show that they refer to redshifts derived from red-sequence colours at the given redshift. The left panel shows data for a slice centred on z$_{phot}=$0.87 and the right a z$_{phot}=$0.96 slice, corresponding to the two peaks in the redshift histogram.  The width of the slices in the cluster-finding algorithm are set by the average colour error around M$^\star$ at each redshift and the width approximately corresponds to $\delta z=0.1$, so there is some overlap between the two model red-sequences\footnote{Thus, it makes negligible difference to the significance contours whether we centre the redshift of the colour slice, z$_{phot}$, on the spectroscopic redshift, z$_{spec}$, or the value given in the RCS catalogue (see Table 3) which is z$_{RCS}$=0.94.}.  This means that some of the same broad structure can be seen in both panels (e.g., near the centre of the field), but that most of the contours in the z$_{phot}=$0.96 slice are of higher significance; i.e., the {\it peak} of the central overdensity occurs around z$_{phot}=$0.96, but the {\it shoulder} of the distribution is still visible in the z$_{phot}=$0.87 slice.  Crosses show galaxies with spectroscopic redshifts incompatible with the redshift of the slice, and squares show galaxies whose redshifts are compatible with being at the redshift of the slice.  The widths of the slices used for the {\it spectroscopic} redshifts is 5000\kms~rest-frame in order to encompass all of the structure visible in both components of the higher redshift system shown in Fig.~3.  It is immediately apparent that the galaxies in the z$_{spec}$=0.96 slice are spatially concentrated within the z$_{phot}$=0.96 contours, thus confirming that our association of this peak in the redshift histogram with our cluster candidate is valid.

Peaks which are identified as cluster candidates are labelled on both maps with their redshift, z$_{RCS}$\footnote{We label these as z$_{RCS}$ to emphasise that they represent a ${\it peak}$ (i.e., a cluster candidate) as opposed to structure of arbitrary significance at the redshift z$_{phot}$. } and significance in parentheses.  The aforementioned z$_{RCS}$=0.94 cluster is the most significant peak in the whole field with $\sigma_{RCS}$=4.7.  

We now consider the identity of the z$_{spec}$=0.87 peak in the redshift histogram.  A 3.1-$\sigma_{RCS}$ cluster candidate appears in the RCS catalogue at z$=$0.88, 4 arc minutes west of the field centre.  Recall that our limit for the final catalogue is $\sigma_{RCS}$=3.3.  This system has 3 spectroscopic members (from the z$_{spec}=$0.87 slice) located within the contours shown and a further two members just outside.  Thus, by our earlier criterion, this would be considered a confirmed cluster, except for the fact that it lies at a lower significance than clusters in the final RCS catalogue, and even lower than the $\sigma_{RCS}$=3.2 clusters (RCS034850-1017.6 and RCS110708-0355.3) considered confirmed.  Regardless, it is clear from these two plots that the red-sequence technique has correctly disentangled the z$_{RCS}$=0.94 peak from any z$_{phot}$=0.88 structure, and in fact correctly identify the z$_{spec}$=0.869 structure as a low significance cluster..

\subsubsection{RCS033414-2824.6}
RCS033414-2824.6 (z$=$0.668) was also observed with LDSS-2 as part of the survey of
Blindert et al.\ (in prep, hereafter B07). They observed masks at three positions
around the cluster: a central pointing close to the position used in
this paper, plus north and south flanking fields. Their redshift
catalogue adds 18 secure cluster members within the region covered by our data.  We note in passing
that there are four objects in common between our surveys.  For only
one of these do we both measure a redshift. Our redshift measurements
of this galaxy, a cluster member, agree to within 70 \kms. This is a
useful independent check of our measurements as Blindert et al.\  used
completely different reduction software and redshift measuring code.

\subsubsection{RCS034850-1017.6}
RCS034850-1017.6 is the only cluster for which an overdensity in redshift space could not
be identified.  Fig.~2 shows a paucity of galaxies above z$\sim$0.8,
the redshift for the cluster estimated from the RCS method.  It is
possible that the depth of the spectroscopy was insufficient to
identify galaxies at z$\gsim$0.8.

In order to test this possibility, we used redshifts from the other
cluster fields and sampled them, mimicking the selection function from
the RCS034850-1017.6 observations in the following way.  We selected
every galaxy for which a redshift was successfully measured from all
the fields except RCS034850-1017.6 and excluded galaxies within 15000
\kms~of the cluster redshift in each.  This formed our field
distribution.  We added galaxies from one of the confirmed z$\sim$0.8
clusters. This formed our mock cluster field. Next we chose galaxies
with redshifts from the RCS034850-1017.6 field, and formed a histogram of
their magnitudes in 0.5 magnitude bins.  This gave the magnitude
selection function: the number of galaxies as a function of magnitude
for which a redshift could be obtained.  We randomly sampled galaxies
from our mock cluster field by applying this selection function (with
Poissonian errors on the number drawn from each bin), and examined the
redshift histogram of the resulting simulated observation, as in
Fig.~2, and applied the techniques described in \S3.2 to see if we
identified the cluster.

In 1000 repeated bootstraps of this method, we failed to identify the
cluster in any realisation. This result is unchanged using either of
the z$\sim$0.8 clusters (RCS110634-0408.9 at z=0.823, or
RCS110708-0355.3 at z=0.825).  We conclude that, at the 3-$\sigma$
level, we could not have identified a z=0.8 cluster in the LDSS-2
spectroscopy if one was present.  We note that if we were to repeat
this test for the z=0.723 cluster, RCS110246-0426.9, then we would
identify an overdensity 14\% of the time; but the z=0.735 cluster,
RCS110723-0523.2, is not identified in any of the 1000 realisations.
So, a z$\sim$0.70 cluster could be marginally detected if one was
present, but a z$\sim0.73$ one would not. 
Thus, our failure to confirm a cluster
based on these data does not necessarily represent a false positive in the RCS
method, but rather is consistent with the limitations of our spectroscopic data.

\subsubsection{RCS141658+5305.2}
This field shows three peaks in the redshift histogram over the whole GMOS field.  Fig.~5 shows the spatial distribution of galaxies in each of these peaks.  Clearly, the galaxies in the middle peak, z$\sim$0.89, are more spatially concentrated than galaxies in the other peaks.  Indeed, the lowest redshift peak, z$\sim$0.61, does not appear at all concentrated and galaxies are spread across the entire GMOS field.  This peak is best interpreted as large scale structure rather than a cluster.  The image is centred on the position of the cluster candidate and so it can be seen that not only are the members of the z$\sim$0.89 peak spatially concentrated, but they are also concentrated around the position of the cluster candidate.  The highest redshift peak, z$\sim$0.97 also appears somewhat concentrated around this area, but not to the extent of the z$\sim$0.89 galaxies.  The object identified as the brightest cluster galaxy in the image is a member of the z$\sim$0.891 peak, as are many of its brightest neighbours.  Thus we associate this peak with the RCS cluster candidate, even though the z$\sim$0.97 peak more closely matches the predicted redshift of the cluster from the RCS technique.

\begin{figure}
\epsscale{0.8}
\plotone{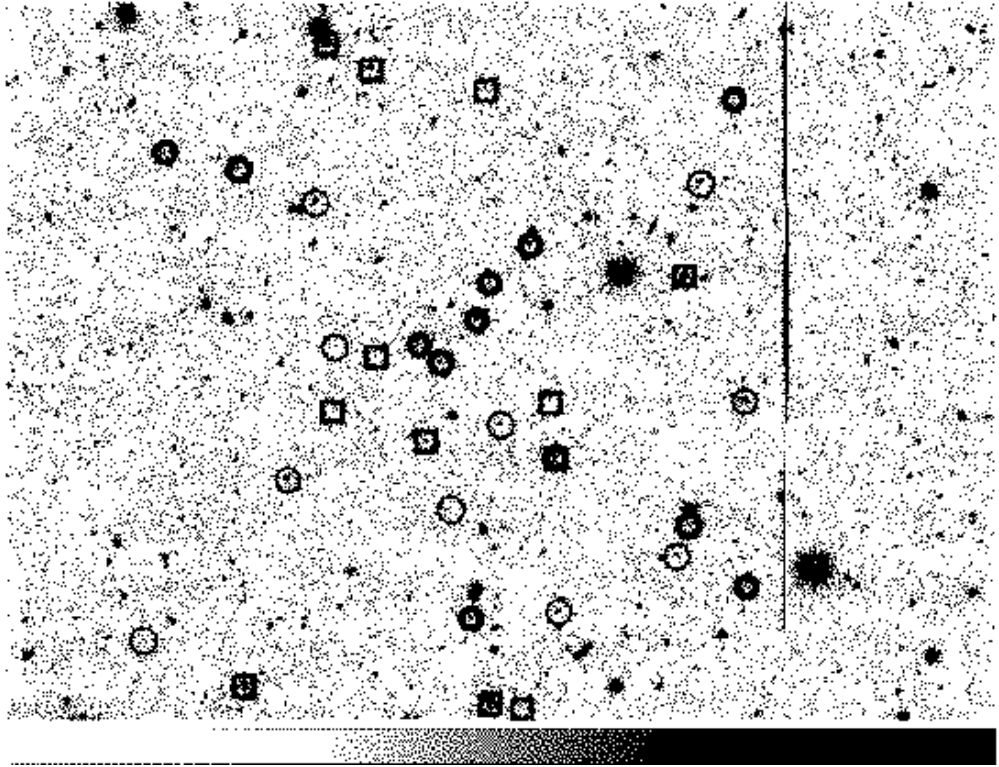}
\caption{RCS141658+5305.2 with galaxies labelled according to spectroscopic redshift.  Objects associated with the z$\sim$0.61 peak are denoted by large, thin circles; z$\sim$0.89 thick circles; z$\sim$0.97 thick squares.  The image is centred on the position of the cluster candidate from the RCS catalogue.  The z$\sim$0.89 galaxies are clearly concentrated around the position of the cluster candidate.  The z$\sim$0.61 galaxies are distributed across the field and so represent large scale structure rather than a genuine cluster.  Image is 6.5 $\times$ 5.0 arcmins, corresponding to 3.0 $\times$ 2.0 $h^{-1}$Mpc at a redshift of 0.89.}
\end{figure}

\subsubsection{Other clusters}

\citet{cop99,ram00,2004MNRAS.348..551G} have all argued that finding 3
galaxies within a velocity range appropriate for that of a cluster's
velocity dispersion is significant.  If we adopt this criterion, all
our clusters (except RCS034850-1017.6) would be significant
detections. The lowest quality redshifts needed for this confirmation
are class 3.  We have demonstrated that even our lowest quality redshift
flag (class 4) is reproducible between different datasets.  Furthermore,
empirical evidence from IMACS follow-up of RCS043938-2904.8 is very
suggestive that these detections based on 3 or 4 redshifts will be
supported by deeper spectroscopy.

It can be seen from Fig.~2 that the highest redshift candidates exhibit
far fewer redshifts than those of the lower redshift clusters.  We are
clearly approaching the limit for measuring redshifts in the optical
with LDSS-2. The decreased sensitivity in the red of this instrument is
such that even the prominent features such as the Ca{\sc[ii]} H \& K are
not readily identifiable at z$\sim$1.

\subsection{Accuracy of the estimated cluster redshifts}

A comparison of the photometric estimates of the cluster redshifts with
those of the measured spectroscopic redshifts is shown in Fig.~6.  The photometric redshifts shown in the figure are raw photometric redshifts from the previous generation of cluster-finding.  These are based purely on population synthesis models for the evolution of the red-sequence.  The cluster-finding method involves a recalibration step \citep[as described in][]{gy00} to empirically bring the average model colours into agreement with the observed colours, as a function of redshift for a subsample of the clusters with spectroscopy.  A low-order polynomial is fitted to photometric vs spectroscopic redshift and the photometric redshifts re-evaluated to minimise the offset.  We emphasise that this correction is applied to {\it all} photometric redshifts and there is no correction of redshifts on an individual cluster basis.  Since the redshift data presented in this paper have been used as part of the correction, it is more instructive to examine how well the redshift estimation technique works before applying this correction.  The dashed line indicates the one-to-one  relation.  The data show a slight trend to overestimate the true
redshift of the cluster at the highest redshift end.  The best fit
relation is indicated by the dotted line and given by $z_{\rm spec}=(0.88\pm0.05)
z_{\rm RCS} + (0.05\pm0.04)$.  The average bias in the redshift
estimate (e.g., \citealt{wittman}) is $\Delta z/(1+z_s)=(0.039\pm0.035)$.  
It is important to note that the scatter in Fig.~6 (i.e., before this correction) is small.  This shows that even using only model colours for the red-sequence, the photo-z estimate is very good, and improved further by a small correction.  The correction just minimises the average offset between photometric and spectroscopic redshifts, as a function of photometric redshift.  The corrected values of the photometric redshift, as used in the latest RCS catalogue, are given in Table~1.  These values give a final accuracy of the red-sequence redshift estimate in this redshift range of 10\%.

\begin{figure}
\plotone{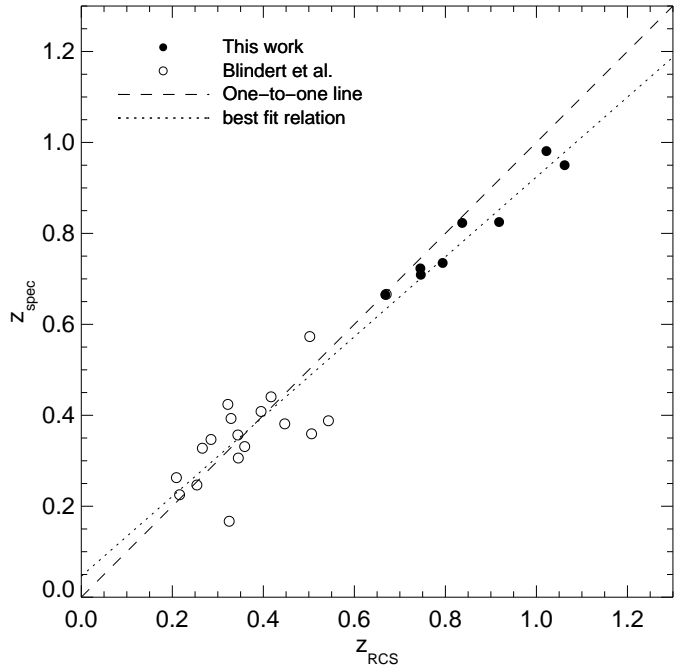}
\caption{Comparison of red-sequence redshift estimates with the measured spectroscopic redshifts of the clusters.  Filled points are from this work and open points are from B07.  The dashed line indicates the one-to-one relation and the dotted line is the best fit.  There is a slight tendency for the 
initial photo-z based on model colours to overestimate the redshift at the high redshift end, but note that the scatter in the relation is small, indicating that a simple rescaling of the estimated redshifts will improve the final accuracy.}
\end{figure}

\subsection{Richness estimates of velocity dispersion}

Four systems yielded sufficient members to attempt to calculate velocity dispersions (see \S3.2) for the clusters: RCS033414-2824.6, RCS035231-1020.7, RCS110723-0523.3 and RCS162009+2929.4.  
For these clusters, we can compare the measured values of the velocity dispersions with the values implied using the relation of \cite{Yee:2003we} from CNOC1 clusters. 
Fig.~7 shows their B$_{gc}$ vs velocity dispersion. Also shown on the plot are data from low-redshift (z$<$0.6) RCS-1 clusters (B07).  The moderately high-redshift RCS-1 clusters presented here appear consistent with both the relation for lower redshift X-ray clusters and the lower redshift relation for RCS-1 clusters, with the exception of the outlier RCS033414-2824.6.  This cluster
has a much lower measured velocity dispersion than inferred from its richness.  Using the additional redshift data from B07 does not change the measured value of the velocity dispersion, within the measurement errors. We interpret this high richness, low velocity dispersion system as a much less massive system, i.e., a group, embedded in richer surrounding sheet-like large scale structure.  Indeed, the unusual sheet-like nature of this structure is clear from the wider-field spectroscopy of B07.

\begin{figure}
\plotone{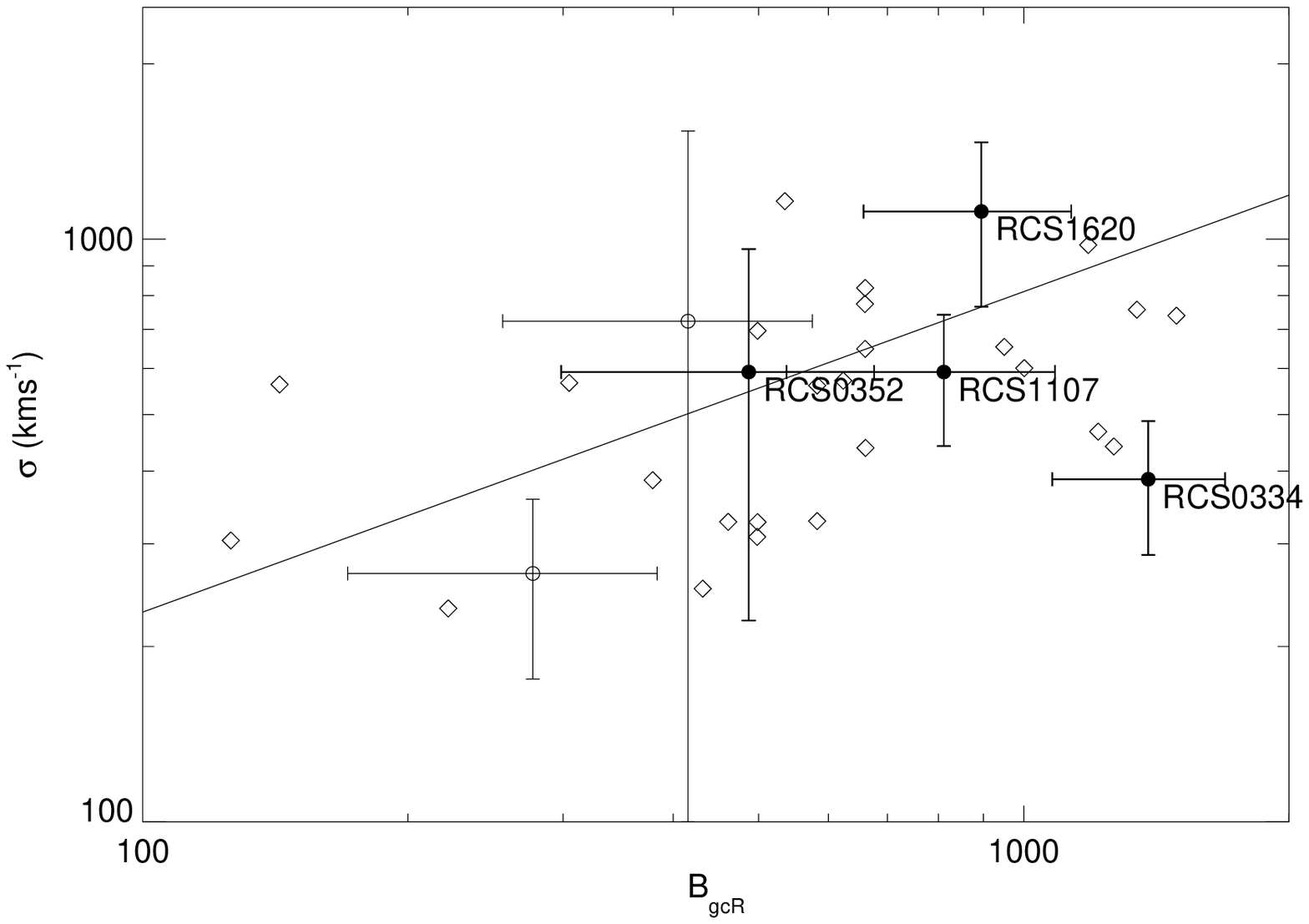}
\caption{Velocity dispersion vs richness estimate (red-sequence \bgc).  Bold points with error bars show the four clusters  (labelled) with sufficient members to consider velocity dispersions reliable.  Open points with error bars are for the two systems in RCS141658+5305.2 with the \bgc~estimate corrected to divide the richness between the two clusters.    Line indicates the relation from \citet{Yee:2003we} derived from the X-ray selected CNOC1 clusters.  Open diamonds (error bars omitted for clarity) show data points from low-redshift RCS-1 (B07).  The moderately high-redshift RCS-1 clusters presented here appear consistent, within the broad uncertainties, with both the relation for X-ray clusters and the lower redshift relation for RCS clusters, with the exception of RCS033414-2824.6. See text for further discussion.  Error bars represent jackknife uncertainties only and do not include any potential systematic error which may be present with modest numbers of redshifts (see \S3.2 for discussion).}
\end{figure}

Such outliers to this relation are expected, but, in the following, we argue as to why we might expect them to be rarer than finding one in this current, modest sample might suggest.
The earliest cluster candidates for spectroscopic follow-up observations (including RCS033414-2824.6) were prioritised using visual inspection, before accurate \bgc~estimates had been calculated.  This may lead to a bias toward selecting systems embedded in sheet-like structures (i.e., low velocity dispersion outliers like RCS033414-2824.6) for a given \bgc.  Consider two cluster candidates with equivalent \bgc s.  This means an equal overdensity of red galaxies (relative to a fixed global background) within 0.5 $h_{50}^{-1}$ Mpc.  Now, if the better candidate of these two is to be selected by eye, initially the eye checks for a concentration of galaxies within some relatively small radius (which we have just set to be the same for both, by construction).  After this, preference is likely to be given to the cluster which has the greatest overdensity on larger scales, since the eye cannot impose a strict cut-off in radius, as the \bgc-measuring algorithm does.  Thus, a system embedded in surrounding structure is likely to look more impressive and be given higher priority for follow-up than a cluster of comparable \bgc~not embedded in larger structure.  This is particularly true if overdensities of red galaxies are searched for using colour pictures, as was the case with some of the early follow-up selection.

Furthermore, we note that RCS033414-2824.6 is also the most distant outlier in the \bgc--$\sigma$ relation from the sample of $\sim$30 clusters of B07.  Thus we expect to find lower incidences of such extreme outliers in the ongoing RCS spectroscopic follow-up, selected using cuts in \bgc~and not relying on visual inspection.  

It is possible that objects like this, embedded in sheet-like structure, could be identified by comparing \bgc~values measured at several different radii.  We are investigating methods involving using another parameter, such as \bgc~concentration, to try to identify potential outliers like RCS033414-2824.6 from the survey data alone.

Similarly, the projection of two or more clusters may cause objects to fall off this relation; e.g., RCS043938-2904.2 comprises two distinct systems in redshift, but close enough that the two may be physically associated.  Thus, direct application of the Virial Theorem to estimate a mass from the velocity dispersion would not be valid. The richness measured for this system would be the sum of the richnesses of the two systems and thus should not be expected to correlate with its mass.  
RCS141658+5305.2 also comprises two systems projected along the line of sight, but separated sufficiently in redshift that the two systems are unrelated and thus velocity dispersions may be calculated individually for both systems.  However, the red-sequences are so close together in colour, $\delta(R-z^\prime)=0.07$, that red-sequence richness estimates for each cluster are contaminated by galaxies from the other.  In order to correct for this, we recalculate the \bgc~values by dividing the measured values between each system in proportion to the numbers of spectroscopic members in each.  These are shown as the two open points with error bars.  It can be seen that, after this correction, the two clusters lie on the relation, albeit with large errors due to the small number ($<$10) of redshifts going into each velocity dispersion estimate.  

Without spectroscopy, using the survey data alone, we would not be able to identify such systems as projections.  However, it should be noted that these projection effects (both physically associated projections/multi-component clusters, and unrelated line-of-sight projections) are present with all cluster-finding methods: e.g., in the X-ray selected CNOC1 sample \citep{Yee:1996lk}, one of the 15 MS clusters (MS0906+11) was found to be binary from the detailed spectroscopy.  Similarly, unrelated systems along the same line of sight are also seen projected in X-ray selected surveys, but examples are relatively scarce in the literature, due to the need for extensive spectroscopic follow-up to reveal such situations.  Several examples of unrelated projected systems at low-redshift for an X-ray luminous sample of Abell clusters are given in \citet{Lopez-Cruz:2004ie}. The main advantage of X-ray selection is that the mass varies less steeply as a function of X-ray luminosity than optical richness.  Thus, projecting two similarly massive clusters together gives a smaller boost to the X-ray luminosity (and hence the detectability in an X-ray survey) than to the optical richness.  To measure the frequency of projections within the RCS requires larger spectroscopic samples and such work is ongoing.  One might expect the projection rate to increase toward the high redshift end of the sample, where red-sequences for different redshifts become degenerate (z$\gsim$0.8 for this filter set).  We currently lack the data to test the redshift dependence of the projection rate within the RCS.  However, an initial estimate can be made by adding the 10 confirmed clusters studied here to those of B07.
Of the 19 RCS systems they studied at 0.3$<$z$<$0.6\footnote{The z$>$0.3 cut is used to avoid the degeneracy of the $(R-z^\prime)$ colour slices at the low redshift end}, B07 find only one comparably-rich system whose red-sequence actually appears to be made up of the projection of a pair of equally rich clusters.  If we adopt redshift bins of 0.3$<$z$<$0.8 and 0.8$<$z$<$1.0, we then find that the fraction of projected systems is 1/23 and 2/6 respectively.  Assuming Poisson errors leads to rates of (4$\pm$4)\% and (33$\pm$24)\%.  Thus, there is slight evidence ($\sim$1$-$sigma) to suggest that the projection rate may increase at z$>$0.8.  A detailed analysis of the expected projection rate derived from cosmological simulations will be presented in future work.  

The concordance of the points from our moderately high-redshift sample with the lower-redshift \bgc--$\sigma$ relation is in good agreement with the results from the cosmological study of \cite{Gladders:2006pp} who found that, based on a self-calibration technique, the evolution in the mass -- \bgc~relation over the redshift range 0.35 to 0.95 was consistent with no evolution.  We note that the definition of \bgc~includes a passively-evolving luminosity limit for the galaxies included in the measurement, so a result of no evolution in this relation means that evolution in the mass--richness relation is consistent with simple passive evolution of the red-sequence cluster galaxies. \cite{Lin:2006cf} and \cite{Muzzin:2006io} also recently found that the evolution between cluster mass and total $K$-band galaxy number (or luminosity) is consistent with passive evolution of the member galaxies.

\section{Conclusions}

We have performed multi-object spectroscopy of 11 RCS clusters at
moderately high redshifts (z$\sim0.7-1.0$).  Using a very conservative
criterion we clearly confirm 7 of the 11 clusters.  Another 3 are
confirmed using the less stringent requirement of 3 galaxies within 1500
\kms~of each other.  Deeper spectroscopy of one of these 3 clusters
supports the reality of this system, and we use this to argue that 10 of
the 11 systems should be considered confirmed clusters.  We demonstrate
that for the remaining cluster candidate the spectroscopic data are too
shallow to have identified the cluster, and that this does not
necessarily constitute a false positive in the RCS technique.  In addition this
cluster lies just below the significance threshold for the final
cluster catalogue and would not have been included.  
While a much larger sample of both clusters and redshifts is needed to quantitatively assess the contamination rate as a function of cluster redshift, these first results are broadly consistent with the $\sim$5\% false-positive rate stimated from simulations \citep{Gladders:2002ui}.  

The RCS technique provides redshift estimates accurate to within 10\% in this redshift range.
Two of the RCS clusters comprise projections of pairs of comparably rich systems.  
In one of these, the two components are close enough in redshift that they may be physically related.  Thus we might consider these two projections to be made up of a) a binary cluster (RCS043938-2904.4; such a binarity fraction would be comparable, within the large uncertainties for such a small sample, with that found in other cluster surveys); and b) an artificially enhanced detection due to the projection of two unrelated clusters (RCS141658+5305.2).  We note that, in both of these cases, the clusters lie at z$>$0.8 and this may be due to the increasing degeneracy of red-sequence colours at these redshifts.  In the former case, we demonstrate how the red-sequence technique also reliably disentangles the cluster from foreground structure.  Comparison with the sample of B07 supports the idea that the projection rate from unrelated clusters at lower redshifts in the RCS survey is likely to be lower.

We present a first look at the correlation between cluster richness, \bgc, and velocity dispersion for a subsample of six clusters at these redshifts.  These measurements appear consistent, within broad uncertaities, with the relation found at lower redshift.

This paper presents initial results from a larger campaign of follow-up
spectroscopy of moderately high and high-redshift clusters from the Red-Sequence Cluster Survey.  A high-redshift sample based on observations with  FORS2 on the VLT and GMOS on Gemini will be reported by \citet{felipe07}.  An ambitious project using ultraplex IMACS observations at the Magellan Baade telescope of a well-defined core sample from the RCS, targeting $\sim$40 clusters selected in bins of richness and redshift (covering 0.3$<$z$\lsim$0.85) is underway.  A spectroscopic survey of a comparable number of RCS clusters, extending the high redshift end to z$\sim$1 using the upgrade to LDSS-2, LDSS-3, is ongoing.  

\section*{acknowledgements}
We are grateful to Dan Kelson and Bob Abraham for making available their
excellent spectroscopy software and for discussions regarding its use and Gus Oemler for providing the COSMOS software.  

The RCS project is supported by grants to H.K.C.Y. from the Canada Research Chair Program,  the Natural Sciences and Engineering Research Council of Canada (NSERC) and the University of Toronto.

EE acknowledges NSF grant  AST-0206154.

M.D.G. acknowledges previous support from NSERC via a post-doctoral
fellowship. Partial support for this work was provided by NASA through
Hubble Fellowship grant HF-01184.01 awarded by the Space Telescope Science
Institute, which is operated by the Association of Universities for
Research in Astronomy, Inc., for NASA, under contract NAS 5-26555.

LFB acknowledges the support of the FONDAP center for Astrophysics and CONICYT under proyecto  FONDECYT 1040423
\\


\begin{thebibliography}{30}

\bibitem[{{Abraham} {et~al.}(2004){Abraham}, {Glazebrook}, {McCarthy},
  {Crampton}, {Murowinski}, {J{\o}rgensen}, {Roth}, {Hook}, {Savaglio}, {Chen},
  {Marzke}, \& {Carlberg}}]{2004AJ....127.2455A}
{Abraham}, R.~G., et al. 2004, \aj, 127, 2455

\bibitem[{{Allington-Smith} {et~al.}(1994){Allington-Smith}, {Breare}, {Ellis},
  {Gellatly}, {Glazebrook}, {Jorden}, {Maclean}, {Oates}, {Shaw}, {Tanvir},
  {Taylor}, {Taylor}, {Webster}, \& {Worswick}}]{Allington-Smith:1994qp}
{Allington-Smith}, J., et al. 1994, \pasp,
  106, 983

\bibitem[{{Barrientos} {et~al.}(2004){Barrientos}, {Gladders}, {Yee},
  {Infante}, {Ellingson}, {Hall}, \& {Hertling}}]{2004ApJ...617L..17B}
{Barrientos}, L.~F., {Gladders}, M.~D., {Yee}, H.~K.~C., {Infante}, L.,
  {Ellingson}, E., {Hall}, P.~B., \& {Hertling}, G. 2004, \apjl, 617, L17

\bibitem[{{Barrientos} {et~al.}(2007){Barrientos}, {Gilbank}, {Gladders},
  {Yee}, {Infante}, {Ellingson}, {Hall}, \& {Hertling}}]{felipe07}
{Barrientos}, L.~F., {Gilbank}, D.~G., {Gladders}, M.~D., {Yee}, H.~K.~C.,
  {Infante}, L., {Ellingson}, E., {Hall}, P.~B., \& {Hertling}, G. 2007, in
  prep

\bibitem[{{Beers} {et~al.}(1990){Beers}, {Flynn}, \& {Gebhardt}}]{bfg90}
{Beers}, T.~C., {Flynn}, K., \& {Gebhardt}, K. 1990, \aj, 100, 32

\bibitem[{{Bigelow} {et~al.}(1998){Bigelow}, {Dressler}, {Shectman}, \&
  {Epps}}]{Bigelow:1998fj}
{Bigelow}, B.~C., {Dressler}, A.~M., {Shectman}, S.~A., \& {Epps}, H.~W. 1998,
  in Proc. SPIE Vol. 3355, p. 225-231, Optical Astronomical Instrumentation,
  Sandro D'Odorico; Ed., ed. S.~{D'Odorico}, 225--231

\bibitem[{{Blakeslee} {et~al.}(2006){Blakeslee}, {Holden}, {Franx}, {Rosati},
  {Bouwens}, {Demarco}, {Ford}, {Homeier}, {Illingworth}, {Jee}, {Mei},
  {Menanteau}, {Meurer}, {Postman}, \& {Tran}}]{Blakeslee:2006bd}
{Blakeslee}, J.~P., et al. 2006, \apj, 644, 30

\bibitem[{{Blindert} {et~al.}(2007){Blindert}, {Yee}, {Gladders}, {Ellingson},
  {Gilbank}, {Barrientos}, \& {Golding}}]{Blindert:07prep}
{Blindert}, K., {Yee}, H.~K.~C., {Gladders}, M.~D., {Ellingson}, E., {Gilbank},
  D.~G., {Barrientos}, L.~F., \& {Golding}, J. 2007, in prep
  
\bibitem[{{Danese} {et~al.}(1980){Danese}, {de Zotti}, \& {di
  Tullio}}]{Danese:1980ng}
{Danese}, L., {de Zotti}, G., \& {di Tullio}, G. 1980, \aap, 82, 322

\bibitem[{{Ford} {et~al.}(2004){Ford}, {Postman}, {Blakeslee}, {Demarco},
  {Jee}, {Rosati}, {Holden}, {Homeier}, {Illingworth}, \&
  {White}}]{2004astro.ph..8165F}
{Ford}, H., et al. 2004, ArXiv Astrophysics e-prints

\bibitem[{{Gal} \& {Lubin}(2004)}]{Gal:2004gt}
{Gal}, R.~R. \& {Lubin}, L.~M. 2004, \apjl, 607, L1

\bibitem[{{Gilbank} {et~al.}(2004){Gilbank}, {Bower}, {Castander}, \&
  {Ziegler}}]{2004MNRAS.348..551G}
{Gilbank}, D.~G., {Bower}, R.~G., {Castander}, F.~J., \& {Ziegler}, B.~L. 2004,
  \mnras, 348, 551

\bibitem[{{Gladders}(2002)}]{Gladders:2002ui}
{Gladders}, M.~D. 2002, Ph.D.~Thesis

\bibitem[{{Gladders} \& {Yee}(2000)}]{gy00}
{Gladders}, M.~D. \& {Yee}, H.~K.~C. 2000, \aj, 120, 2148

\bibitem[{{Gladders} \& {Yee}(2005)}]{Gladders:2005oi}
---. 2005, \apjs, 157, 1

\bibitem[{{Gladders} {et~al.}(2006){Gladders}, {Yee}, {Majumdar}, {Barrientos},
  {Hoekstra}, {Hall}, \& {Infante}}]{Gladders:2006pp}
{Gladders}, M.~D., {Yee}, H.~K.~C., {Majumdar}, S., {Barrientos}, L.~F.,
  {Hoekstra}, H., {Hall}, P.~B., \& {Infante}, L. 2006, astro-ph/0603588

\bibitem[{{Glazebrook} \& {Bland-Hawthorn}(2001)}]{2001PASP..113..197G}
{Glazebrook}, K. \& {Bland-Hawthorn}, J. 2001, \pasp, 113, 197

\bibitem[{{Gunn} {et~al.}(1986){Gunn}, {Hoessel}, \& {Oke}}]{gho86}
{Gunn}, J.~E., {Hoessel}, J.~G., \& {Oke}, J.~B. 1986, \apj, 306, 30

\bibitem[{{Holden} {et~al.}(1999){Holden}, {Nichol}, {Romer}, {Metevier},
  {Postman}, {Ulmer}, \& {Lubin}}]{cop99}
{Holden}, B.~P., {Nichol}, R.~C., {Romer}, A.~K., {Metevier}, A., {Postman},
  M., {Ulmer}, M.~P., \& {Lubin}, L.~M. 1999, \aj, 118, 2002

\bibitem[{{Kelson} {et~al.}(2000){Kelson}, {Illingworth}, {van Dokkum}, \&
  {Franx}}]{2000ApJ...531..159K}
{Kelson}, D.~D., {Illingworth}, G.~D., {van Dokkum}, P.~G., \& {Franx}, M.
  2000, \apj, 531, 159

\bibitem[{{Kurtz} \& {Mink}(1998)}]{1998PASP..110..934K}
{Kurtz}, M.~J. \& {Mink}, D.~J. 1998, \pasp, 110, 934

\bibitem[{{Levine} {et~al.}(2002){Levine}, {Schulz}, \&
  {White}}]{2002ApJ...577..569L}
{Levine}, E.~S., {Schulz}, A.~E., \& {White}, M. 2002, \apj, 577, 569

\bibitem[{{Lima} \& {Hu}(2004)}]{Lima:2004jt}
{Lima}, M. \& {Hu}, W. 2004, \prd, 70, 043504

\bibitem[{{Lin} {et~al.}(2006){Lin}, {Mohr}, {Gonzalez}, \&
  {Stanford}}]{Lin:2006cf}
{Lin}, Y.-T., {Mohr}, J.~J., {Gonzalez}, A.~H., \& {Stanford}, S.~A. 2006,
  astro-ph/0609169

\bibitem[{{L{\'o}pez-Cruz} {et~al.}(2004){L{\'o}pez-Cruz}, {Barkhouse}, \&
  {Yee}}]{Lopez-Cruz:2004ie}
{L{\'o}pez-Cruz}, O., {Barkhouse}, W.~A., \& {Yee}, H.~K.~C. 2004, \apj, 614,
  679
  
\bibitem[{{Muzzin} {et~al.}(2006){Muzzin}, {Yee}, {Hall}, {Ellingson}, \&
  {Lin}}]{Muzzin:2006io}
{Muzzin}, A., {Yee}, H.~K.~C., {Hall}, P.~B., {Ellingson}, E., \& {Lin}, H.
  2006, astro-ph/0612202

\bibitem[{{Ramella} {et~al.}(2000){Ramella}, {Biviano}, {Boschin}, {Bardelli},
  {Scodeggio}, {Borgani}, {Benoist}, {da Costa}, {Girardi}, {Nonino}, \&
  {Olsen}}]{ram00}
{Ramella}, M., {Biviano}, A., {Boschin}, W., {Bardelli}, S., {Scodeggio}, M.,
  {Borgani}, S., {Benoist}, C., {da Costa}, L., {Girardi}, M., {Nonino}, M., \&
  {Olsen}, L.~F. 2000, \aap, 360, 861

\bibitem[{{Rosati} {et~al.}(1999){Rosati}, {Stanford}, {Eisenhardt}, {Elston},
  {Spinrad}, {Stern}, \& {Dey}}]{1999AJ....118...76R}
{Rosati}, P., {Stanford}, S.~A., {Eisenhardt}, P.~R., {Elston}, R., {Spinrad},
  H., {Stern}, D., \& {Dey}, A. 1999, \aj, 118, 76

\bibitem[{{Sarazin}(2002)}]{Sarazin:2002rv}
{Sarazin}, C.~L. 2002, {The Physics of Cluster Mergers} (ASSL Vol.~272: Merging
  Processes in Galaxy Clusters), 1--38

\bibitem[{{Vanden Berk} {et~al.}(2001){Vanden Berk}, {Richards}, {Bauer},
  {Strauss}, {Schneider}, {Heckman}, {York}, {Hall}, {Fan}, {Knapp},
  {Anderson}, {Annis}, {Bahcall}, {Bernardi}, {Briggs}, {Brinkmann}, {Brunner},
  {Burles}, {Carey}, {Castander}, {Connolly}, {Crocker}, {Csabai}, {Doi},
  {Finkbeiner}, {Friedman}, {Frieman}, {Fukugita}, {Gunn}, {Hennessy},
  {Ivezi{\'c}}, {Kent}, {Kunszt}, {Lamb}, {Leger}, {Long}, {Loveday}, {Lupton},
  {Meiksin}, {Merelli}, {Munn}, {Newberg}, {Newcomb}, {Nichol}, {Owen}, {Pier},
  {Pope}, {Rockosi}, {Schlegel}, {Siegmund}, {Smee}, {Snir}, {Stoughton},
  {Stubbs}, {SubbaRao}, {Szalay}, {Szokoly}, {Tremonti}, {Uomoto}, {Waddell},
  {Yanny}, \& {Zheng}}]{2001AJ....122..549V}
{Vanden Berk}, D.~E., et al. 2001, \aj, 122, 549

\bibitem[{{Wittman} {et~al.}(2001){Wittman}, {Tyson}, {Margoniner}, {Cohen}, \&
  {Dell'Antonio}}]{wittman}
{Wittman}, D., {Tyson}, J.~A., {Margoniner}, V.~E., {Cohen}, J.~G., \&
  {Dell'Antonio}, I.~P. 2001, \apjl, 557, L89

\bibitem[{{Yee} \& {Ellingson}(2003)}]{Yee:2003we}
{Yee}, H.~K.~C. \& {Ellingson}, E. 2003, \apj, 585, 215

\bibitem[{{Yee} {et~al.}(1996){Yee}, {Ellingson}, \& {Carlberg}}]{Yee:1996lk}
{Yee}, H.~K.~C., {Ellingson}, E., \& {Carlberg}, R.~G. 1996, \apjs, 102, 269

\end{thebibliography}
\end{document}